# Prediction of polarization vortices, charge modulation, flat bands, and moiré magnetism in twisted oxide bilayers

Naafis Ahnaf Shahed[1], Kartik Samanta[1], Mohamed Elekhtiar[1], Kai Huang[1],
Chang-Beom Eom[2], Mark S. Rzchowski[3], Kirill D. Belashchenko[1], and Evgeny Y. Tsymbal[1]

[1] *Department of Physics and Astronomy & Nebraska Center for Materials and Nanoscience,
University of Nebraska, Lincoln, Nebraska 68588-0299, USA*

[2] *Department of Materials Science and Engineering, University of Wisconsin, Madison, Wisconsin 53706, USA*

[3] *Department of Physics, University of Wisconsin-Madison, Madison, Wisconsin 53706, USA*

The recent surge of interest in moiré superlattices of twisted van der Waals compounds has spotlighted the emergence of unconventional superconductivity and novel electronic phases. However, the range of moiré phenomena can be dramatically expanded by incorporating complex oxide materials into twisted heterostructures. In this study, motivated by the recent breakthroughs in synthesis of free-standing oxide membranes, we explore the emergent structural and electronic properties of twisted oxide bilayers. We focus on the classic perovskite oxide, $SrTiO_3$, and design $SrTiO_3$ bilayers with a relative twist between the individual layers. Using density functional theory calculations, we predict the appearance of vortex-antivortex polarization patterns at the interface of the $SrTiO_3$ bilayers driven by twist. We also predict charge modulation of the interfacial Ti ions induced by varying local coordination which follow the moiré pattern. Furthermore, we forecast the emergence of flat bands at large twist angles and the associated localized electronic states with moiré-periodic charge density, originating from the interlayer bonding effects resulting in the formation of dangling bonds. Finally, we predict that hole doping induces unconventional $d^0$ magnetism in otherwise nonmagnetic $SrTiO_3$, driven by the exchange splitting of the high-density O-$p$ bands and producing the spin density with moiré periodicity. These results demonstrate a broad landscape of emergent phenomena which may occur in moiré-engineered oxide heterostructures showing far-reaching perspectives of these material systems for further fundamental studies and potential applications.

## I. INTRODUCTION

Theoretical predictions of nearly dispersionless (flat) bands with vanishing electronic velocity in a twisted bilayer graphene [1,2] have aroused significant interest in the scientific community and led to establishing a new field of research dubbed twistable electronics or twistronics [3-7]. Twistronics involves non-trivial properties and phenomena of artificial moiré superlattices emerging in two-dimensional (2D) materials like graphene due to a relative twist angle between successive layers. The hallmark of twistronics has been the observation of unconventional superconductivity driven by a flat band in a graphene bilayer twisted at the "magic" angle of about 1.1° [8]. This discovery triggered extraordinary efforts to search for other materials with similar electronic behavior. Flat bands have been found in twisted bilayers made of other homogeneous materials, such as transition metal dichalcogenides [9-12], hexagonal BN [13,14], $C_3N$ [15], $In_2Se_3$ [16], GeSe [17], and $BC_3$ [18], as well as in bilayer [19] and trilayer [20-24] heterostructures. In addition to superconductivity, a periodic moiré potential and orbital overlap, as well as lattice reconstruction between sites of different atomic registry, give rise to other non-trivial quantum phases, which are not intrinsic to the parent materials. For example, moiré superlattices are predicted to generate orbital ferromagnetism [25-28], Wigner crystallization [29], quantum anomalous Hall effects [30], topological behavior [31-33], unconventional ferroelectricity [34], exitonic insulator phases [35], and novel photonic and non-linear optical properties [36,37].

The expected and observed novel properties of moiré superlattices are often driven by strong electron-electron correlations. The flat bands are characterized by a large effective mass of electrons and, as a result, by a greatly suppressed electron's kinetic energy. This causes electrons to be localized in space and be affected by strong electron-electron interactions. Destructive interference of electronic waves scattered by different lattice sites has been considered as a prerequisite for the flat band formation trapping the electron within a localized region [38,39]. Destructive interference usually occurs when two bonds in a unit cell share a common site, commonly seen in triangular lattice structures like the Kagome lattice [40], the side-centered square lattice (the Lieb lattice), [41,42] and the checkerboard lattice. [43] Twisting one layer with respect to the other in a multilayer allows exploiting the features of these special lattices to observe the strong localization of electrons.

So far, most efforts to observe novel properties of moiré superlattices have been focused on 2D van der Waals (vdW) compounds where relatively weak interlayer interactions facilitate their mechanical exfoliation. However, the range of moiré phenomena can be dramatically expanded by incorporating complex oxide materials into twisted heterostructures. The strongly interacting electrons of the *d* orbitals in transition-metal oxides give rise to additional rich spectrum of striking phases, such as high-temperature superconductivity, colossal magnetoresistance, Mott metal-insulator transitions, and multiferroicity. Furthermore, a long wavelength modulating potential in twisted vdW materials is relatively weak, producing energy band splitting typically of a few meV. Such small energy splitting makes these weakly localized electronic states prone to thermal fluctuations and are destroyed at room temperature. The exchange coupling between layers is also relatively weak, acting as a perturbation on the intralayer spin and electronic properties. In contrast, the interlayer modulation potential in oxide moiré heterostructures can be significantly enhanced due to much stronger interlayer interactions. The larger band splitting is expected to make flat electronic bands robust to thermal fluctuations and disorder. The interfacial atomic rearrangements can lead to strain-driven polarization vertices and topological defects. Oxide crystals are stable in ambient conditions, providing robustness often lacking in vdW heterostructures.

Synthesis and assembly of moiré superlattices with these complex oxide materials is more challenging compared to their vdW counterparts. Unlike conventional vdW materials that can be easily exfoliated down to a monolayer without disrupting the structural stability, oxides cannot be exfoliated into pure 2D layers due to strong bonding between the layers. Recently, however, two techniques have been developed to synthesize freestanding oxide membranes. A widespread method of synthesizing oxide membranes is the epitaxial growth of water-soluble layer on perovskite substrates, followed by *in-situ* growth of oxide films and heterostructures. [44,45] Another method employs the "physical lift-off" technique, which allows synthesizing freestanding oxide membranes with different crystallographic orientations. [ 46 ] The ultrathin oxide membranes exhibit different behavior from their bulk counterparts. For example, ultrathin $SrTiO_3$ and $BiFeO_3$ freestanding layers display exceptional elasticity despite being ceramic materials that are typically brittle in their bulk form. [45] The fabrication of heterostructures using remote epitaxy is also possible by stacking these freestanding layers of oxides on top one another. [46] Furthermore, the precise control of the twist angle between oxide membranes has been demonstrated, [47-49] which lays the foundation for oxide twistronics. New polar, topological, and magnetic phases are expected to emerge in these oxide moiré superlattices. For example, stacking $BaTiO_3$ membranes with controlled twist angles has revealed a pattern of polarization vortices and antivortices driven by the strain gradient. [49]

The goal of this study is to demonstrate a broad landscape of emergent phenomena which may occur in moiré-engineered oxide heterostructures. We employ density functional theory (DFT) and tight-binding models to investigate the structural and electronic properties of twisted oxide heterostructures. Our focus is on a representative perovskite oxide, $SrTiO_3$, known for its stable atomic structure, which is non-conducting, non-polar, and non-magnetic in a pristine bulk form. Using extensive DFT calculations, we demonstrate that the assembly of $SrTiO_3$ thin films in a twisted bilayer structure forms a moiré superlattice strongly affecting structural, electronic, and magnetic properties of $SrTiO_3$. Specifically, we observe the formation of the vortex-antivortex displacement patterns, suggesting an unconventional polar phase arising from twist. We predict the emergence of charge modulation on the interfacial Ti ions which form the moiré superlattice. We forecast the appearance of ultra-flat electronic bands occurring at relatively large twist angles and exhibiting charge localization with moiré periodicity. We also demonstrate unconventional $d^0$ magnetism driven by the spin splitting of the O-*p* bands in response to hole doping, resulting in moiré-periodic spin density. Overall, our results suggest promising avenues for moiré-engineered oxide heterostructures to reveal new properties that are not present in conventional oxide thin-film structures and may be interesting from the fundamental physics point of view and useful for novel applications.

## II. ATOMIC STRUCTURE OF A TWISTED $SrTiO_3$ BILAYER

Our target material, $SrTiO_3$, has a perovskite structure (space group Pm-3m) and can be thought of as (001) monolayers of SrO and $TiO_2$ stacked upon one another along the [001] direction [Fig. 1(a)]. To maintain the chemical and structural integrity, each $SrTiO_3$ layer in a twisted bilayer should contain at least one unit cell of bulk $SrTiO_3$ to preserve the chemical bonds of the perovskite structure. Therefore, in our calculations, we assume that the bottom $SrTiO_3$ layer has two unit cells of $SrTiO_3$ and is terminated with the SrO monolayer on bottom and with the $TiO_2$ monolayer on top. The top $SrTiO_3$ layer has one and a half unit cells and terminated on both sides with the SrO monolayers. The resulting twisted $SrTiO_3$ bilayer contains seven monolayers and preserves the conventional stacking of bulk perovskite $SrTiO_3$. The bilayer has $SrO/TiO_2$ interface and the same SrO termination on top and bottom sides [Figs. 1(b) and 1(c)].

To build a twisted supercell with commensurate structures of the two $SrTiO_3$ layers, we use the concept of Pythagorean triples $(l, m, n)$, such that $n^2 = l^2 + m^2$. This allows us to find commensurate twist angles given by $\theta = sin^{-1}(l/n)$. In this work, we build four twisted $SrTiO_3$ structures with Pythagorean triples: (3, 4, 5), (5, 12, 13), (8, 15, 17), and (7, 24, 25), corresponding to twist angles $\theta$ of 36.9°, 22.6°, 28.1°, and 16.3°, respectively, and moiré supercell lengths of 8.63 Å, 15.9 Å, 13.9 Å, and 19.3 Å, respectively.

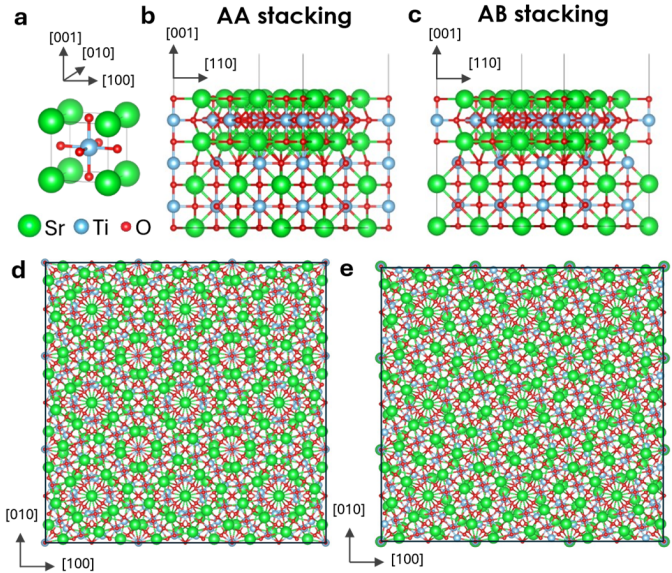

**Fig. 1** Atomic structure of twisted SrTiO$_3$. (a) Conventional unit cell of perovskite SrTiO$_3$. (b, c) Side view of the 22.6°-twisted SrTiO$_3$ supercell with AA (b) and AB (c) stacking orders along the twisting [001] axes (indicated). (d, e) Top view of the 3×3 cell of the 22.6° twisted SrTiO$_3$ bilayer with AA (d) and AB (e) stacking, showing the formation of moiré patterns.

The lateral alignment of the twisted SrTiO$_3$ layers has freedom of in-plane sliding of the top layer with respect to the bottom. We explore two limiting cases: (1) *AA stacking*, where the bulk-like Ti-O-Ti-O chain is preserved across the twisted interface at the corner of the moiré supercell [Fig. 1(b)], and (2) *AB stacking*, where O-Ti-O chain in the top twisted layer breaks the bulk sequence across the interface and is continuation of the Sr-Sr atomic chain of the bottom layer at the corner of the moiré supercell [Fig. 1(c)]. Note that for both stacking orders, the conventional TiO$_2$-SrO sequence of the SrTiO$_3$ (001) monolayers is preserved across the interface. Within the AA stacking, the Sr-Sr chain continues across the interface at the center of the moiré supercell and the Ti-Ti chain continues at the corners. On the contrary, within the AB stacking, the Sr-Sr sequence in the top twisted layer is continuation of the Ti-O-Ti atomic chain of the bottom layer at the center of the moiré supercell. The AB stacking is obtained from the AA stacking by shifting the top twisted layer by half a unit cell in the [110] direction. As seen from Figures 1(d) and 1(e), both structures exhibit pronounced moiré patterns in the plane of the bilayer but have different in-plane geometry.

The atomic relaxation of the bilayer structures is performed using DFT calculations as described in Appendix A. During relaxation, we observe a notable restructuring of the interface. In both stacking configurations, there is a tendency for the twisted layers to seek a larger separation as compared to that in bulk. In the AB-stacked bilayer with a twist angle $\theta$ = 22.6°, the Sr-O bond length at the interface (which is preserved in the center of the moiré supercell for this stacking) expands to 1.99 Å, marking a 0.06 Å increase from the respective bulk bond length. In contrast, the AA-stacked bilayer does not have an Sr-O bond at the interface. To measure the extent of separation at the twisted interface, we analyze the Ti-O bond in the upper twisted layer at the supercell corner adjacent to the interface and find it to be 1.78 Å. This value is notably smaller than the bulk Ti-O bond length of 1.93 Å, indicating an enlarged interfacial gap for this stacking arrangement as well. Despite this increase, the interlayer separation at the interface remains relatively modest compared to typical vdW materials. [50-52] This fact indicates that the twist operation does not break all bonds across the interface but changes them according to interface geometry.

Our calculations predict that the twisted structures with AA stacking have lower energy than AB stacking independent of the twist angle. For example, in the case of $\theta$ = 22.6°, we find that the energy difference is about 23 meV per surface unit cell of bulk SrTiO$_3$. While this energy difference is nonvanishing, we argue that the AB stackings forming a metastable state can be realized experimentally, due to a relatively large lateral shift ~2.8Å that is required to transform it to the AA stacking. For a large moiré cell, this shift is energetically prohibitive due to strong local bonds between the atoms across the interface.

Regardless of the degree of twist and stacking order, we consistently observe reconstruction of the Ti-O bonds within the TiO$_2$ layers adjacent to the twisted interface. Interestingly, some of these bonds are broken leading to reduced coordination of interfacial Ti atoms and, as a result, to their reduced oxidation state. This leads to Ti ionic charge modulation which has moiré periodicity, as discussed in Sec. IV.

In addition, we find that displacement of Ti atoms with respect to their unrelaxed positions forms a vortex-like pattern. Figure 2(a) shows the atomic structure of the TiO$_2$ monolayer nearest to the interface in the top (twisted) AA-stacked SrTiO$_3$ layer. The vortices have circular- or cylindrical-like shapes and exhibit alternating chirality. The vortices are interleaved with antivortices, both forming a periodic pattern with moiré periodicity. Similar displacements are observed in the bottom SrTiO$_3$ layer; however, the resulting vortices there have an opposite chirality compared to those in the top layer [Fig. 2(e)]. Such vortex-like displacement patterns have been predicted previously for non-vdW twisted layers like PbS [53] and experimentally demonstrated in perovskite oxide ferroelectric BaTiO$_3$. [49] They are somewhat reminiscent of those observed in polar untwisted oxide superlattices. [54] In complex oxide materials, this leads to polarization vortex patterns, as described in Sec. III.

Similar features are found in an AB-stacked SrTiO$_3$ bilayer with twist angle $\theta$ = 22.6° (see Appendix B for details). Altering the twist angle changes the moiré supercell size and thus periodicity of the vortex-like patterns.

## III. STRAIN-INDUCED POLARIZATION VORTICES

The calculated Ti displacement patterns in the plane of $TiO_2$ monolayers represent distributions of the intralayer strain that is spatially varied with the same periodicity as the moiré superlattice, demonstrating strong interaction between the two twisted $SrTiO_3$ layers. To quantify these strain patterns, we follow Sánchez-Santolino *et al.* [49] and calculate the shear strain that is defined by

$$\epsilon_{xy} = \epsilon_{yx} = \frac{1}{2}\left(\frac{\partial u_y}{\partial x} + \frac{\partial u_x}{\partial y}\right),$$

where $u_x$ and $u_y$ are displacements of the Ti atoms along the $x$ and $y$ directions with respect to their positions in the unrelaxed structures. The results shown in Figures 2(b) and 2(f) indicate that shear strain maps in the twisted AA-stacked $SrTiO_3$ bilayer reveal alternating regions of positive and negative $\epsilon_{xy}$. Shear strain appears to be nearly a factor of 3 stronger for the $TiO_2$ monolayer in the bottom portion of the stack [Fig. 2(f)] compared to the $TiO_2$ monolayer in the top portion of the stack [Fig. 2(b)]. This is due to the former lying exactly at the interface, contrary to the latter being next to the interfacial SrO monolayer. It is notable that in the same lateral regions, the shear strain in the two $TiO_2$ monolayers have opposite signs [compare Figs. 2(b) and 2(f)]. We find that shear strain maps of the twisted AB-stacked $SrTiO_3$ bilayer reveal similar features (see Appendix B for details).

Due to Ti atoms being charged in an ionic-type compound like $SrTiO_3$, the observed Ti atom displacements are polar, hence representing polarization patterns in $TiO_2$ monolayers adjacent to the interface. These polarization patterns are expected to correlate with the strain. Bulk $SrTiO_3$ is not, however, piezoelectric, and hence a linear relationship between polarization and strain is not expected. We note though that even in ferroelectric $BaTiO_3$, the linear piezoelectric coupling between polarization and strain does not play the dominant role. [49] We therefore focus on the in-plane strain gradients in the interfacial $TiO_2$ monolayers, which are expected to produce polarization through flexoelectric coupling. [55]

Our analysis of the calculated strain gradients along with the vortex-like polarization patterns supports the flexoelectric coupling mechanism in twisted $SrTiO_3$ bilayers. Specifically, for an AA-stacked $SrTiO_3$ bilayer, there are regions in the top (twisted) portion of the bilayer with positive and negative strain gradients $\partial\varepsilon_{xy}/\partial x$ represented as green and brown regions in Figure 2(c). It is evident that these regions display positive and negative displacements of the Ti atom in the $TiO_2$ monolayer along the $y$ axis, respectively [light gray arrows in Fig. 2(c)]. Consequently, the $y$-component of polarization $P_y$ is directly correlated with the shear strain gradient. This correlation is mirrored in the relationship between the strain gradient $\partial\varepsilon_{xy}/\partial y$ and the $x$-component of polarization $P_x$ [Fig. 2(d)].

Qualitatively similar features of the strain gradients are observed in the $TiO_2$ monolayer in the bottom portion of the stack [Figs. 2(g) and 2(h)]. We see again correlation between strain gradients $\partial\varepsilon_{xy}/\partial x$, $\partial\varepsilon_{xy}/\partial y$ and polarization components $P_y$, $P_x$, respectively. Reflecting the enhanced strain, strain gradients are a factor of 3 stronger in the interfacial bottom $TiO_2$ monolayer compared to the top. Also, it seen that the color maps of $\partial \epsilon_{xy}/\partial x$ and $\partial \epsilon_{xy}/\partial y$ are interchanged between the two TiO$_2$ monolayers [compare Figs. 2(c,d) and 2(g,h)], reflecting an opposite sign of polar displacements in the top and bottom SrTiO$_3$ layers. Similar correlation between the polarization components and strain gradients are observed in the AB-stacked bilayer (Appendix B and Fig. A1).

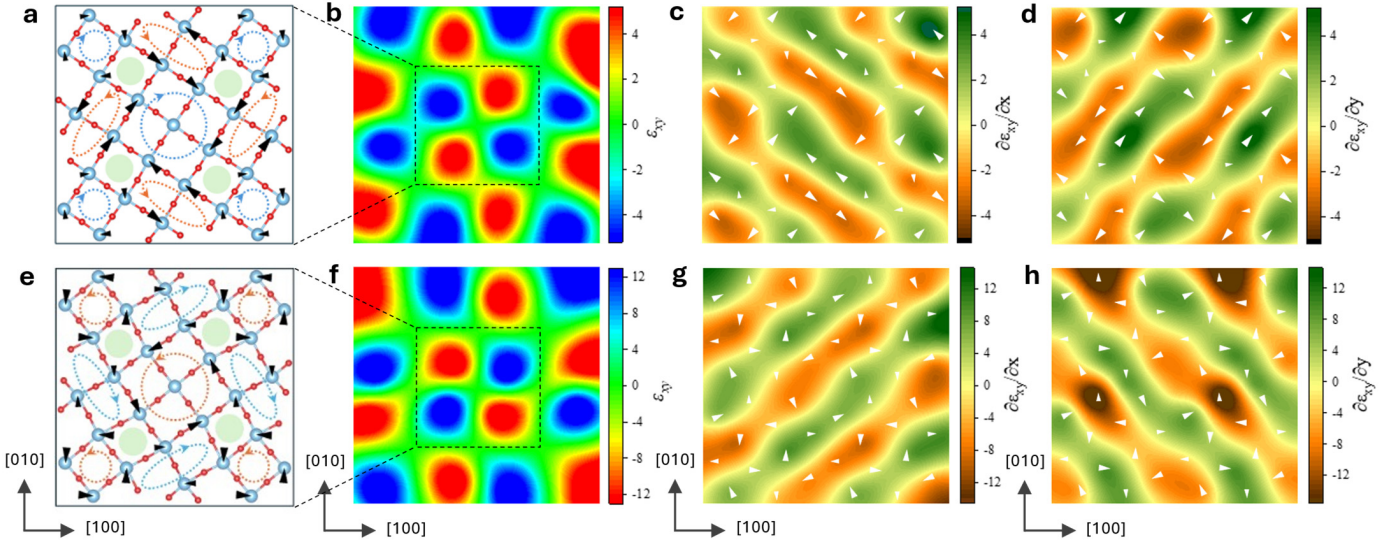

**Fig. 2** Formation of polarization vortices in a 22.6° twisted $SrTiO_3$ bilayer with AA stacking. (a, e) Atomic structure of the $TiO_2$ monolayer nearest to the interface in the top (a) and bottom (e) $SrTiO_3$ layers. Black arrows (proportional in size) show displacement of the Ti atoms with respect to their unrelaxed positions. Dashed circles and ellipses (differed in color and arrow direction) indicate the formation of displacement vortices of opposite chirality. Regions highlighted in green signify antivortices. (b-c, f-h) Shear strain (b, f) and shear strain gradient (c-d, g-f) maps obtained for $TiO_2$ monolayers near the interface in the top (b,c,d) and bottom (f,g,h) $SrTiO_3$ layers. Gray arrows (proportional in size) show displacement of the Ti atoms with respect to their unrelaxed positions.

All these features indicate that there is a linear relationship between polarization and strain gradient which can be described as follows [49]

$$P_x = \mu_{xyxy} \frac{\partial \epsilon_{xy}}{\partial y}, \qquad P_y = \mu_{xyxy} \frac{\partial \epsilon_{xy}}{\partial x}, \qquad (1)$$

where $\mu_{xyxy}$ is an effective flexoelectric coefficient. We conclude therefore that the vortex-like polarization patterns emerge due to the flexoelectric coupling in both TiO$_2$ monolayers adjacent to the interface. It is notable that while SrTiO$_3$ is non-polar in the bulk form, when twisted, it exhibits qualitatively similar polarization features as those found in twisted ferroelectric BaTiO$_3$. [49]

### IV. BOND AND CHARGE MODULATION

Twisting an oxide bilayer leads to the interfacial atomic reconstruction changing electronic properties of the twisted interface compared to the bulk compound. We find that the atomic registry of the interfacial Ti atoms varies continuously within the moiré superlattice between three distinct types of atomic configurations. In a twisted AA-stacked SrTiO$_3$ bilayer these configurations are linked to the three commensurate lattice sites (CLSs) where Ti, Sr, or O atomic chains are continuous across the interface [Fig. 3(a)]. As seen from Figure 3(a), while at the Ti- and Sr-CLSs the Ti atoms have the 6-fold coordination like those in bulk SrTiO$_3$, at the O-CLS the incommensurability of the Ti atoms across the interface breaks out-of-plane Ti-O bonding leading to the reduced 5-fold coordination of the Ti atoms. Figure 3(b) illustrates the local atomic structure around the O-CLS at the interface of a twisted AA-stacked SrTiO$_3$ bilayer. It is seen that the O atoms within the top SrO monolayer have broken bonds with the Ti atoms within the TiO$_2$ monolayer resulting in the 5-fold coordination of these interfacial Ti atoms.

The observed bond-order reconstruction results in the electronic reconstruction of the twisted SrTiO$_3$ interface which is reflected in the changing oxidation state of the interfacial Ti atoms. We quantify this behavior by calculating Density-Derived Electrostatic and Chemical (DDEC) charges on the Ti atoms [56], as detailed in Appendix C. The DDEC method uses smart partitioning of the weighted electron density to obtain chemically meaningful net atomic charges and allows quantifying charge transfer between atoms involving ionic and covalent bonding and dielectric screening.

To correlate the calculated DDEC charges on the interfacial Ti atoms in a twisted SrTiO$_3$ bilayer with their conventional oxidation states, we calibrate the DDEC charges against the Ti nominal ionic charges in several titanium oxides and bulk SrTiO$_3$, as demonstrated in Figure A2(a). Then, using this calibration, we estimate the nominal oxidation state of the interfacial Ti atoms. We find that these Ti atoms are largely divided into two categories: those which are located at the Ti- or Sr-CLSs and have an oxidation state of about 3.9+ that is close to 4+ as in bulk SrTiO$_3$ and those which at located at the O-CLSs and have the oxidation state of about 3.5+ [Fig. 3(b)]. The reduced oxidation state of the Ti atoms at the O-CLSs is due to the broken Ti-O bond, leading to their 5-fold coordination rather than 6-fold coordination like in bulk SrTiO$_3$. The excess electron charge on the Ti atoms due to the broken bond reduces their oxidation state. Similar results are obtained based on the Bader charge analysis, [57] showing that our conclusions are qualitatively independent of the charge density partitioning to designate the net atomic charges (Appendix C). The predicted charge modulation of the interfacial Ti atoms exhibits moiré periodicity.

To obtain further insight into the charge modulation, we calculate the electronic density of states (DOS) projected onto the 3$d$-orbitals of the interfacial Ti atoms. Figure 3(e) indicates that the weight of the Ti-3$d$ states at the O-CLS Ti atoms is higher and shifted to low energies [red curve in Fig. 3(e)] as compared to that on the other Ti atoms [grey area in Fig. 3(e)]. By integrating the partial DOS and scaling the obtained number of electrons per Ti atom to the total number of valence electrons, we find that the O-CLS Ti atoms accommodate approximately 0.3$e$ per atom more than the non-O-CLS Ti atoms and Ti atoms in bulk SrTiO$_3$. This result suggests that the oxidation state of the O-CLS Ti atoms is reduced by about 0.3$e$, which is in qualitative agreement with our findings based on the DDEC and Bader charge analyses.

The partial DOS in Figure 3(e) suggests that the largest contribution to the excess charge on the O-CLS Ti sites comes from the DOS peak at the lowest energies of around -4.5 eV. This fact indicates that the excess electron charge due to the broken Ti-O bond participates in the covalent bonding of the O-CLS Ti to another O atom and contributes to the strongly bound state. From the charge density calculated in the energy window from -4.7 eV to -4.3 eV, we observe localization of the excess charge along the Ti-O bond with the bottom apex O atom participating in this bonding [Fig. 3(f)]. The strong bonding between these two atoms is also evident from the significantly reduced Ti-O bond length of 1.73 Å compared to the bulk Ti-O bond length of 1.93 Å.

A twisted SrTiO$_3$ bilayer with AB stacking exhibits a qualitatively similar bond and charge modulation behavior, although the details are different. Specifically, we find that the atomic registry of the interfacial Ti atoms exhibits 5- and 6-fold coordination, depending on whether these Ti atoms are located at the Sr/Ti-CLS or Ti/Sr-CLSs [Fig. 3(c)]. The atomic structure of the Sr/Ti-CLS [left panel in Fig. 3(c)] has a Sr atom within the top SrO monolayer lying atop the Ti atom in the interfacial TiO$_2$ monolayer. This configuration breaks the Ti-O bond across the interface so that this Ti atom has 5-fold coordination. In addition, the four nearest Ti atoms reveal a significantly increased bong length up to about 2.43Å with the O atoms in the upper (twisted) SrO monolayer indicating further bond modulation [thin grey Ti-O bond lines in the left panel of Fig. 3(c)]. On the contrary, the atomic structure of the Ti/Sr-CLSs [right panel in Fig. 3(c)] has an O atom within the upper SrO monolayer lying atop an empty site in the lower TiO$_2$ monolayer. This configuration does not break Ti-O bonds across the interface and supports 6-fold coordination of the nearest interfacial Ti atoms.

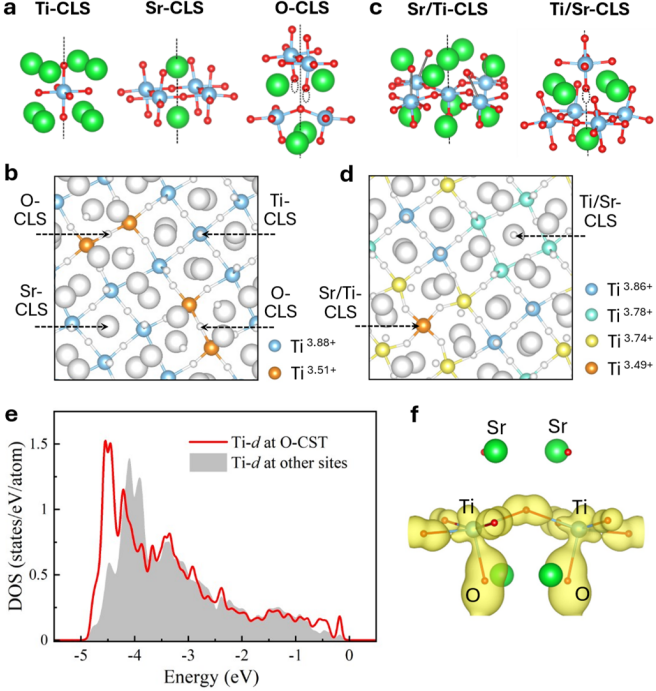

**Fig. 3** Moiré charge modulation in 22.6° twisted SrTiO$_3$ bilayers. (a) Local coordination of the interfacial Ti atoms at the Ti-commensurate lattice site (CLS), and near Sr- and O-CLSs in the AA- stacked bilayer. A dotted ellipse indicates a dangling bond. (b) Interfacial atomic structure of the AA-stacked bilayer indicating positions of the Ti-, Sr-, and O-CLSs and the nominal Ti oxidation states. (c) Local coordination of the interfacial Ti atoms at the Sr/Ti- and Ti/Sr-CLSs in the AB-stacked bilayer. In the left panel, long Ti-O bonds at the four Ti atoms around the Sr/Ti-CLS are shown by thin grey lines. In the right panel, dotted ellipses indicate dangling bonds. (d) Interfacial atomic structure of the AB-stacked bilayer indicating positions of at the Sr/Ti- and Ti/Sr-CLSs and the nominal Ti oxidation states. Sr atom at the Sr/Ti-CLS is removed for clarity. (e) Partial density of states (DOS) projected on the 3$d$ orbitals of the Ti sites at the O-CLSs (red curve) and other interfacial Ti sites (grey area) in AA-stacked bilayer. (f) Isosurface of the charge density in the energy window from -4.7 eV to -4.3 eV at the O-CLS.

Calculating the DDEC on the interfacial Ti atoms and correlating them with their conventional oxidation states of Ti atoms, as described in Appendix C and shown in Fig. A2(b), we observe that the 5-fold coordinated Ti atoms exhibit an oxidation state close to 3.5+, whereas other Ti ions that have 6-fold coordination reveal higher valence states that are closer to that in bulk SrTiO$_3$ [Fig. 3(d)]. The predicted charge modulation of the interfacial Ti atoms exhibits moiré periodicity.

## V. FLAT BANDS

We perform band structure calculations for AA- and AB-stacking orders of the twisted SrTiO$_3$ bilayer as described in Appendix A.1. In Figure 4, we highlight the results of these calculations for $\theta = 22.6°$. For the AA stacking configuration, we observe a band structure that is composed of multiple bands representing conduction and valence bands separated by a band gap of about 1.7 eV [Fig. 4(a)]. Like in bulk SrTiO$_3$, the conduction band is largely composed of the Ti-3$d$ orbitals, while the valence band is majorly composed of the O-2$p$ orbitals [Fig. 4(c)]. The band structure of the AB stacked SrTiO$_3$ bilayer, as seen from Figure 4(b), exhibits, in general, qualitatively similar behavior. However, there is a new feature emerging in the AB stacked SrTiO$_3$ bilayer, that is a band with an exceptionally narrow bandwidth at the top of the valence band. For $\theta = 22.6°$, this band exhibits a bandwidth of just 4.3 meV [inset in Fig. 4(b)], characterizing it as ultra-flat. [12] We see that in contrast to twisted vdW systems, where flat bands typically appear at low twist angles, twisted oxide bilayers can produce flat bands at large twist angles.

The orbital-resolved DOS of the AB-stacked structure [Fig. 4(d)] reveals a sharp peak at the energy corresponding to the flat band, attributing its character to the O-2$p_z$ orbitals and indicating strong electronic localization. This localization is evident from Figure 4(f) which shows the calculated band-resolved charge density distribution in real space associated with the flat band. It is seen that this charge density is localized in the top part of the twisted bilayer around the out-of-plane Ti-O bonds at the corners of the twisted supercell. The AB stacking order effectively isolates these out-of-plane Ti-O bonds within the twisted SrTiO$_3$ layer [Fig. 1(c)] creating a dangling bond [Fig. 1(c), right panel]. As seen from Figure 4(f), there is also an additional charge accumulation around the corners of the twisted supercell at the neighboring oxygen atoms in the TiO$_2$ plane. The localized charge density exhibits moiré periodicity, producing a moiré superlattice of charged dots.

In contrast, the topmost valence band in the AA-stacked SrTiO$_3$ bilayer is dispersive and does not exhibit charge localization. This is seen from Figure A3, demonstrating that the charge density associated with this band is largely delocalized within the interfacial TiO$_2$ monolayer in the bottom part of the SrTiO$_3$ bilayer. There are, however, three bands at around −0.18 eV below the Fermi energy [indicated in red in Fig. 4(a)], which have relatively narrow bandwidth, producing a peak in the DOS at this energy [Fig. 4(c)]. These bands exhibit a higher degree of localization, as seen from the real-space charge density map shown in Figure 4(e). They have, however, a larger bandwidth compared to the bandwidth of the flat bands in the AB-stacked system.

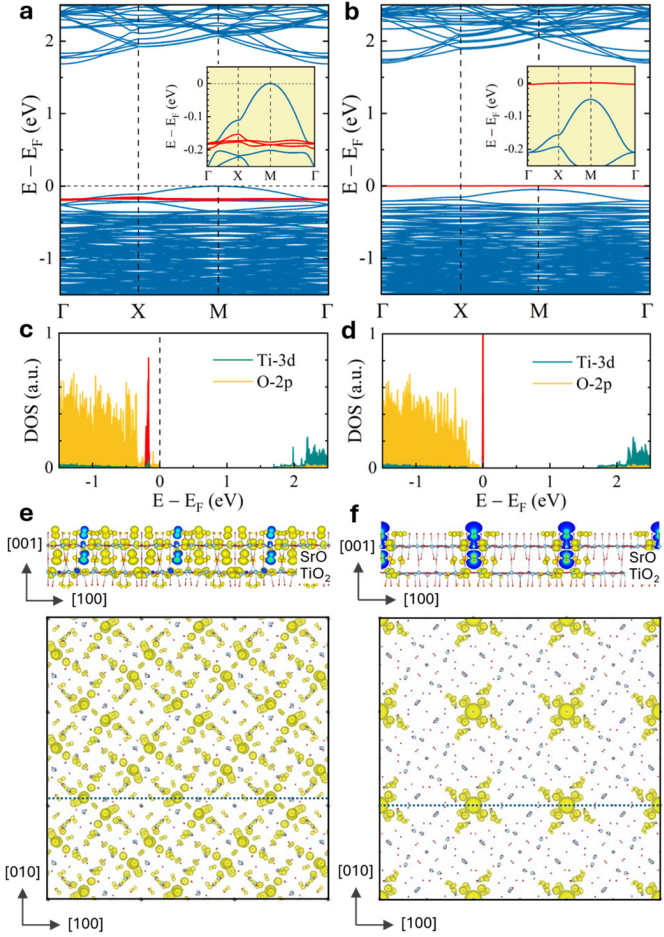

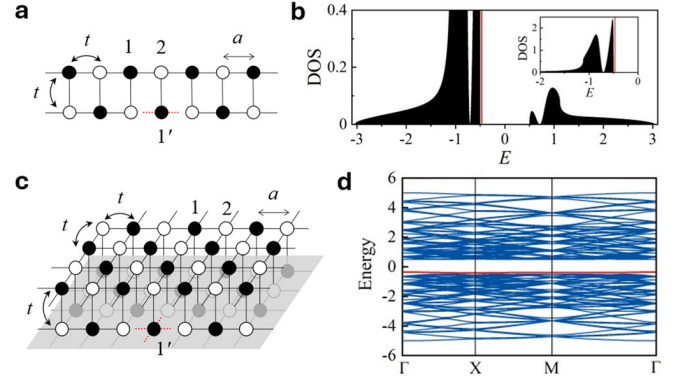

**Fig. 4** Electronic structure and charge density of a 22.6° twisted SrTiO$_3$ bilayer. (a, b) Electronic band structure of AA- (a) and AB- (b) stacked SrTiO$_3$ bilayers along high symmetry lines in the first Brillouin zone. Insets show the band structures in the narrow energy window close to the Fermi energy. Red lines indicate flat bands. (c, d) Orbital-resolved density of states (DOS) of AA- (c) and AB- (d) stacked SrTiO$_3$ bilayers. Red color indicates the DOS produced by the flat bands. (e, f) Isosurface of the charge density (shown in yellow) in real space corresponding to the flat bands in AA- (e) and AB- (f) stacked SrTiO$_3$ bilayers. Top panels: side views, bottom panels: top views. Dotted lines in the bottom panels show the positions of the front planes in the side views. Interfacial SrO and TiO$_2$ monolayers are indicated.

As is evident from Figure 4(e), the large charge density is concentrated at the out-of-plane Ti-O bonds in the top part of the twisted SrTiO$_3$ bilayer. The O atoms participating in this bonding are located above the O-CLSs and have broken bonds with the interfacial Ti atoms [O atoms indicated by dotted ellipses at the O-CLS in Fig. 3(a)]. These dangling bonds are responsible for the formation of flat bands in the AA-stacked SrTiO$_3$ bilayer. Contrary to the AB-stacked bilayer where the flat band lies in the energy gap, the flat bands in the AA-stacked bilayer lie within the continuum of dispersive bands and have a larger bandwidth due to hybridization with the continuum. These bands are reminiscent of the resonant states, which are distinct from the bound states that lie in the band gap.

**Fig. 5** Tight-binding model. (a) Schematics of a 1D model of two coupled atomic chains. Atom 1′ has broken bonds (shown by dashed red lines) with the two nearest atoms in the bottom chain. Red dashed lines show broken bonds. (b) Density of states (DOS) on site 1′ calculated using the Green's function approach for $t = 1$, $\Delta = 0.5$, $\varepsilon_{1'} = -0.6$. The two-band continuum split by a band gap is shown in black and the localized state in red. (c) Schematics of a 2D TB model: a checkerboard atomic lattice of alternating sites 1 and 2. Atom 1′ has broken bonds (indicated by dashed red lines) with four nearest atoms in the plane. (d) Band structure of the model (c) calculated for an 8×8 supercell and $t = 1$, $\Delta = 0.5$, $\varepsilon_{1'} = -0.2$. (a).

Similar trends are observed for the SrTiO$_3$ bilayers with other twist angles. We find flat bands in the AA- and AB-stacked bilayers at all twist angles investigated. At all twist angles, we observe the isolated flat band at the top of the valence band in the AB-stacked bilayers, whereas the flat bands lie always within the continuum states in the AA-stacked bilayers. In all cases, we find that the flat bands appear due to the features of the atomic structure producing dangling bonds above the interface of twisted SrTiO$_3$ bilayer.

To obtain further insight into the origin of flat bands in twisted SrTiO$_3$ structures, we develop a simple tight-binding (TB) model which captures the essential physics of this phenomenon. As noted above, the origin of flat bands is the formation of dangling bonds at the O atoms which are located above the interfacial TiO$_2$ monolayer due to broken bonds with the Ti atoms in this monolayer. The broken bonds appear on a single site in the moiré supercell, while at the other sites, the bonds are unbroken. To reflect this feature in the TB model, we consider two one-dimensional (1D) atomic chains with alternating sites 1 and 2 (mimicking O and Ti atoms, respectively), which have on-site atomic energies, $\varepsilon_{1,2} = \mp\Delta$, and coupled by intra-chain and inter-chain hopping $t$ between nearest-neighbor atoms, as shown in Figure 5(a). We assume that atom $1'$ in the bottom chain forms dangling bonds: it is coupled with hopping $t$ to the nearest atom 2 in the top chain but has broken bonds with the two nearest atoms 2 in the bottom chain [indicated by red dashed lines in Fig. 5(a)]. The atom $1'$ is assumed to have on-site energy $\varepsilon_{1'}$ which may be different from that of atom 1 (i.e., $-\Delta$) reflecting an altered atomic environment of this atom.

Using Green's function formalism, as described in Appendix F, we calculate the local DOS on site $1'$ as a function of energy $E$. The results are shown in Figure 5(b) for $t = 1$, $\Delta = 0.5$, $\varepsilon_{1'} = -0.6$. It is seen that the DOS represents two continuum bands [shown in black in Fig. 5(b)] which are split by a band gap at energies between –0.5 to 0.5. In addition, we see a localized state emerging above the bottom valence-band continuum [highlighted in red in Fig. 5(b)]. This state resembles the localized state found in our DFT calculations for an AB-twisted $SrTiO_3$ bilayer [Fig. 4(a)], indicating that dangling bonds are essential to the formation of the flat band.

The energy and appearance of a localized state depends on the onsite energy $\varepsilon_{1'}$ of atom $1'$. For $\varepsilon_{1'} = -\Delta$, the localized state appears on top of the bottom band. When $\varepsilon_{1'} < -\Delta$, but close to $-\Delta$, a localized state emerges in the band gap, as illustrated in Figure 5(b). In addition, we observe a resonant state in the valence band which is broadened by hybridization with the continuum states (a peak in DOS below $E = -\Delta$). This resembles the broadened localized state found in our DFT calculations for an AA-twisted $SrTiO_3$ bilayer [Fig. 4(b)].

Extending the 1D model to a more realistic 2D TB model (Appendix F), we consider two planer atomic lattices with a checkerboard atomic structure of alternating sites 1 and 2 (mimicking O and Ti atoms, respectively) with on-site atomic energies $\varepsilon_{1,2} = \mp\Delta$ coupled by the nearest-neighbor hopping $t$ within each lattice and between the lattices, as shown in Figure 5(c). We assume that atom $1'$ in the bottom lattice has on-site energy $\varepsilon_{1'}$ and forms dangling bonds: it is coupled with hopping $t$ to the nearest atom 2 in the top lattice but has broken bonds with the four nearest atoms 2 in the bottom lattice [indicated by red dashed lines in Fig. 5(c)]. Atom $1'$ resembles atom 1 (oxygen), whose on-site energy is altered by a different atomic environment. This model replicates a simplified version of the dangling bonds in our twisted structure.

Similar to our 1D model, for $\varepsilon_{1'} = -\Delta$, we find a localized state on top of the valence band (not shown). For $|\varepsilon_{1'}| > \Delta$, the localized state shifts into the valence-band continuum and strongly hybridizes with the continuum states, while for $|\varepsilon_{1'}| < \Delta$, it moves into the band gap, as illustrated in Figure 5(d). Notably, the flat band emerges solely due to the inclusion of dangling bonds in the TB Hamiltonian.

Based on these TB models, we conclude that the appearance of the flat bands is driven by the local atomic structure of the twisted interface. This behavior is different from the effect of destructive interference in the hopping processes typical for the Kagome and Lieb lattices, [40-42] but reminiscent to the mechanism of formation of the impurity states in semiconductors, [58,59] where the impurity potential produces electronic states which have a large density around the impurity center. Such localized states can be of two types: bound states, where their energy lies in a forbidden gap, and resonant states, where the impurity levels lie within the continuum and hybridize with the continuum states. The former are like a flat band at the top of the valence band in the twisted AB-stacked $SrTiO_3$ bilayer [Fig. 4(b)], whereas the latter are like the flat bands overlapped with the dispersive band the twisted AA-stacked $SrTiO_3$ bilayer [Fig. 4(a)].

## VI. MOIRÉ MAGNETISM

The appearance of the valence flat bands in twisted oxide bilayers points out a possibility of unconventional properties such as moiré magnetism which can be activated by proper hole doping. The high density of the O-$p$ states near the Fermi level indicates that the exchange-driven spin splitting of the electronic bands can occur in a nominally nonmagnetic $SrTiO_3$ due to incompletely filled O-$p$ states. This type of $d^0$ magnetism contrasts with conventional magnetism, which typically arises from the partially occupied $d$- or $f$-orbitals. Depending on the specific band structure and relative population of different valence states, the exchange splitting of the spin bands can be mediated by the Stoner mechanism [60] or the Anderson mechanism [61]. According to the Stoner model, spontaneous ferromagnetism occurs when the relative gain in the exchange energy becomes larger than the loss in the kinetic energy, resulting from the spin splitting of electronic bands. In the simplest *single-orbital* magnet, this condition is satisfied when the Stoner criterion is met, i.e., $ID(E_F) > 1$, where $D(E_F)$ is the DOS per spin at the Fermi energy $E_F$ in the nonmagnetic state, and $I$ is the Stoner parameter, reflecting the strength of the exchange interaction [60]. While the Stoner mechanism describes itinerant magnetism, the Anderson mechanism deals with localized impurity states interacting with the continuum of non-spin-polarized free-electron-like bands. In this case, the emergence of local magnetic moments depends on the energy of the (singly filled) impurity level, $\epsilon_{imp}$, relative to the Fermi level, the magnitude of the Hubbard energy at the impurity site $U$, and the width $\delta$ of the virtual state formed by the hybridization of the localized electron with the continuum [61]. The formation of a local moment is favorable when $|\epsilon_{imp}| \gg \delta$ and $\epsilon_{imp} + U \gg \delta$. Interaction between local magnetic moments may, at sufficiently low temperature, lead to the formation of a magnetically ordered state.

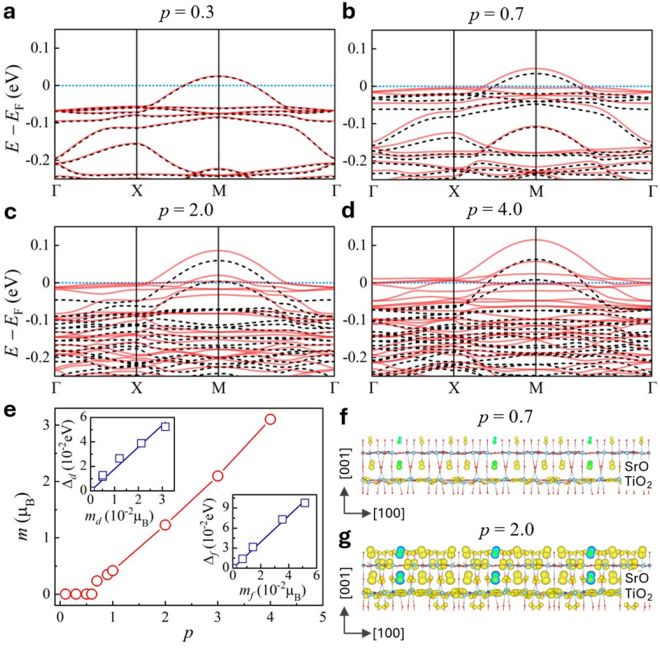

**Fig. 6** Spin-dependent electronic structure, magnetic moment, and spin density of a 22.6°-twisted AA-stacked SrTiO$_3$ bilayer doped with holes. (a-d) Spin-resolved band structure close to the valence band maximum for hole doping concentrations of $p$ = 0.3 (a), 0.7 (b), 2.0 (c), and 4.0 (d) per entire unit cell of twisted SrTiO$_3$. Solid red and dashed black lines indicate minority- and majority-spin bands, respectively. Blue dotted line shows the Fermi energy. (e) Total magnetic moment $m$ as a function of hole doping $p$. Top inset shows the spin splitting $\Delta_d$ of the dispersive top-valence band as a function of the average magnetic moment $m_d$ per O atom at the TiO$_2$ interfacial monolayer (squares) with a linear fit (solid line). Bottom inset shows the spin splitting $\Delta_f$ of the flat band as a function of the average magnetic moment $m_f$ per O atom at the O sites forming the flat band (squares) with a linear fit (solid line). (f, g) Isosurface of the spin density (shown in yellow) for hole doping $p$ = 0.7 (f) and 2.0 (g) given in the same scale. Interfacial SrO and TiO$_2$ monolayers are indicated.

Within the density functional theory (DFT), the stability of a magnetically-ordered state is determined by the Stoner model. Although the resulting criterion may underestimate the stability of local magnetic moments, it is a good approximation in those cases where electron-electron interactions are not too strong. In a *multi-orbital* magnet, the magnetic instability is determined by the generalized Stoner criterion $\det(1 - \hat{\chi}_0 \hat{I}) = 0$, where $\hat{\chi}_0$ is a magnetic susceptibility matrix in the nonmagnetic state, and $\hat{I}$ is the generalized Stoner parameter matrix. The vanishing determinant corresponds to divergent magnetic susceptibility, and the eigenvector of the matrix describes magnetic moments spontaneously appearing on different orbitals.

To simplify this general description, we assume that the relevant electronic bands are formed by the 2$p$ orbitals of two types of O atoms: those which are responsible for the formation of the flat bands [i.e., O atoms forming dangling bonds that are indicated by dotted ellipses at the O-CLS in Fig. 3(a)] and those which shape the dispersive top valence band [O atoms lying within the TiO$_2$ interfacial monolayer where, according to Figure A3(b), the most spin density of this band is located]. Assuming that these orbitals are not strongly hybridized, we find that for small values of spin splitting $\Delta_a$, the Stoner model predicts $\Delta_a = I_a m_a$. Here index $a$ distinguishes the flat bands (denoted by index $a = f$) from the dispersive top-valence band (denoted by index $a = d$), so that the magnetic moment $m_a$ per atom (in units of $\mu_B$), the band splitting $\Delta_a$, and the Stoner exchange parameter $I_a$ are different for the two types of bands.

We then use DFT to study the effect of hole doping on the electronic structure of twisted SrTiO$_3$ bilayers by incrementally removing valence electrons (see Appendix A for details of these calculations). For the twisted AA-stacked SrTiO$_3$ bilayer, we observe that small doping $p$ = 0.3 (here and below the value of $p$ determines the number of holes per unit cell of the twisted structure) does not produce any spin splitting and thus magnetism [Fig. 6(a)]. The spin splitting [Fig. 6(b)] and magnetic moment [Fig. 6(e)] appear at $p$ = 0.7. With the further increase of hole doping, the spin splitting [Fig. 6(c,d)] and the magnetic moment [Fig. 6(e)] monotonically increase.

Figure 6(f) shows the calculated spin density at $p$ = 0.7, indicating that the spin density largely resides within the interfacial TiO$_2$ monolayer and at certain O sites in the SrO twisted monolayer above it. These O sites are located above the O-CLSs and have broken bonds with the interfacial Ti atoms [O atoms indicated by dotted ellipses at the O-CLS in Fig. 3(a)]. As we have already discussed, they form dangling bonds in the top TiO$_2$ monolayer and are responsible for the formation of the flat bands. The spin density within the interfacial TiO$_2$ monolayer appears from the spin splitting of the charge density of the topmost valence band, which is also localized within the interfacial TiO$_2$ monolayer [Fig. A3(d)].

With the increasing hole doping, other spin-split bands become depopulated resulting in their contribution to the net magnetic moment [Fig. 6(b)]. These bands are no longer associated with the interfacial TiO$_2$ monolayer, and, as a result, the charge density becomes more distributed over the whole twisted SrTiO$_3$ bilayer [Fig. 6(h)]. Nevertheless, even at larger doping, as is evident from Fig. 6(h), we observe the enhanced spin density and magnetic moments on the O sites forming dangling bonds. For example, we find that the average magnetic moment on these O sites increases from 0.027 to 0.068 to 0.127 $\mu_B$ when $p$ changes from 1.0 to 2.0 to 4.0, whereas the average magnetic moment on the O atoms at the interfacial TiO$_2$ monolayer increases from 0.005 to 0.012 to 0.031 $\mu_B$. The enhanced magnetic moments have moiré periodicity qualitatively similar to that of the charge density in Figure 4(e). The moiré modulation of the spin density and magnetic moments is the essential feature of magnetism emerging in the hole-doped twisted SrTiO$_3$.

Next, we estimate the Stoner parameter $I_a$ using a linear fit of $\Delta_a$ as a function of the average magnetic moment $m_a$ per O atom at those O sites which dominate in the formation of these bands. For the dispersive top valence band, we average the magnetic moment over O sites within the TiO$_2$ interfacial monolayer and obtain a $\Delta_d$ vs $m_d$ curve that is shown in the top inset of Figure 6(e). From the linear fit, we estimate the Stoner parameter $I_d \approx 1.79$ eV for the top valence band. For the flat bands, we average the magnetic moment over O sites located above the O-CLSs and obtain a $\Delta_f$ vs $m_f$ curve that is shown in the bottom inset of Figure 6(e). From the linear fit, we estimate the Stoner parameter $I_f \approx 1.96$ eV for these flat bands. This value is somewhat different from $I_d$, which is not surprising because the Stoner parameter is not solely determined by a particular type of atom but depends on the spin-dependent exchange-correlation field which alters with the crystal field, chemical environment, bonding, *etc*. In this regard, the estimated values of the Stoner exchange parameter are in good agreement with those obtained for the Hund exchange of the O-$p$ states (ranging from about 1 to 2 eV) using the high-throughput computational analysis. [62]

Then, we employ the Stoner model to elucidate the emergence of magnetism in a twisted AA-stacked SrTiO$_3$ bilayer at $p = 0.7$. Here, we assume that the relevant electronic states are formed by the 2$p$ orbitals of two types of O atoms: 26 O atoms lying within the interfacial TiO$_2$ monolayer and 8 O atoms at the O-CLSs (where, according to Figure 6(f), the most spin density is localized), and that these orbitals develop equal spin splitting at the point of instability. The resulting simplified Stoner criterion then becomes identical to that in a single-orbital case, i.e., $ID(E_F) = 1$, where $I$ is an averaged Stoner parameter for the participating O 2$p$ orbitals, and $D(E_F)$ is the DOS at the Fermi energy $E_F$ in the nonmagnetic state for these orbitals.

Figure A4 shows the evolution of DOS at energies close to the top of the valence band with doping $p$ increasing from 0 to 0.7. As seen, this evolution is different from just a depopulation of the filled states corresponding to the DOS of the undoped bilayer. In addition to moving $E_F$ from the VBM, there is a shift of the high-density flat bands towards the VBM. At $p = 0.7$, where the spin splitting occurs, hole doping places the Fermi energy at about 0.045 eV below the VBM at the edge of the flat band DOS [dashed line in Fig. A4(d)]. At this value of $E_F$, we find that the total non-spin-polarized DOS for the twisted SrTiO$_3$ bilayer is $D_{tot}(E_F) \approx 50.8$ eV$^{-1}$. Dividing by 34 (that is the number of participating O 2$p$ orbitals), we estimate DOS per O atom per spin $D(E_F) \approx 0.75$ eV$^{-1}$. It is therefore evident that for $p = 0.7$, the simplified Stoner criterion, $ID(E_F) > 1$, is satisfied for $I$ between $I_d \approx 1.79$ eV and $I_f \approx 1.96$ eV, leading to spontaneous spin splitting of the bands and the emergence of magnetism in agreement with our DFT calculations. In contrast, for $p = 0.6$, we obtain $D(E_F) \approx 0.46$ eV$^{-1}$, so that $ID(E_F) < 1$ for the given range of $I$, i.e., the Stoner criterion is not satisfied.

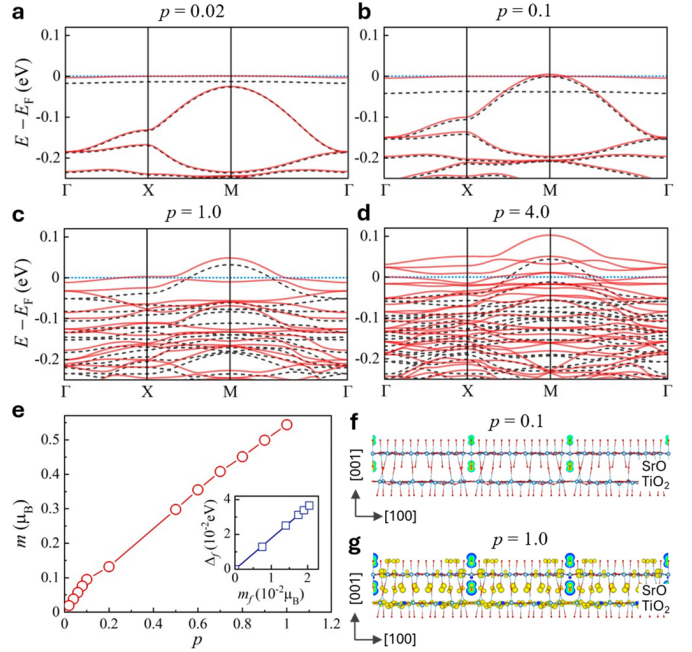

**Fig. 7** Spin-dependent electronic structure, magnetic moment, and spin density of 22.6°-twisted AB-stacked SrTiO$_3$ bilayer doped with holes. (a-d) Spin-resolved band structure close to the valence band maximum for hole doping concentrations of $p = 0.02$ (a), 0.1 (b), 1.0 (c), and 4.0 (d) per entire unit cell of twisted SrTiO$_3$. Solid red and dashed black lines indicate minority- and majority-spin bands, respectively. Blue dotted line shows the Fermi energy. (e) Total magnetic moment $m$ as a function of hole doping $p$. Inset shows the spin splitting $\Delta_f$ of the flat band as a function of the average magnetic moment $m_{af}$ per O atom at the two O sites forming the flat band (squares) with a linear fit (solid line) at small doping $p$. (f, g) Isosurface of the spin density (shown in yellow) for hole doping $p = 0.1$ (f) and 1.0 (g) given in the same scale. Interfacial SrO and TiO$_2$ monolayers are indicated.

For the twisted AB-stacked SrTiO$_3$ bilayer, the situation is different. Due to the presence of the flat band at the Fermi energy, even a very small hole doping leads to its spin splitting [see Fig. 7(a) for $p = 0.02$]. This is because the DOS in the narrow band is so large that the Stoner criterion is satisfied even at a very small doping. The spin-up subband is fully populated, i.e. $n_\uparrow = 1$ [the dashed black line just below $E = E_F$ in Fig. 7(a)], and the spin-down subband is partly empty, i.e. $n_\downarrow = 1 - p$ [the solid red line at $E = E_F$ in Fig. 7(a)], resulting in the total magnetic moment $m = \mu_B p$. According to the Stoner model, $\Delta_f = I_f m_f$, where, in this case, $m_f$ can be estimated from the average moment on the two O atoms forming the out-of-plane Ti-O bonds with the Ti atoms at the top TiO$_2$ monolayer at the corners of the twisted supercell [Figs. 7(f) and 4(f)]. As seen from the inset in Figure 7(e), at small doping, there is a linear increase of the flat band splitting $\Delta_f$ as a function of $m_f$. From this linear variation, we estimate the Stoner exchange constant $I_f \approx 1.78$ eV. This value is in the same range as the values obtained for the AA-twisted SrTiO$_3$.

With the increasing hole doping, the flat band moves to lower energies and its spin splitting gets larger [Fig. 7(b)]. The spin-split flat band overlaps with the dispersive band and hybridizes with it, resulting in its spin splitting as well. At higher doping levels, we observe a complex behavior where the flat band moves deeper into the continuum and becomes strongly hybridized with other bands [Figs. 7(c,d)]. The latter exhibits spin splitting increasing with hole doping. This leads to an increase in the net magnetic moment with doping as shown in Figure 7(e).

Figure 7(f) shows the spin density of the 22.6°-twisted AB-stacked $SrTiO_3$ bilayer for hole doping $p = 0.1$. Notably, that the spin density is localized at the O atoms positioned above and below the Ti-CLSs in the twisted $TiO_2$ monolayer. This is qualitatively similar to the charge density of the flat band [Fig. 4(f)]. Both are largely associated with the localized O-$p_z$ orbitals and have moiré periodicity. At $p = 0.1$, the total magnetic moment on the two O atoms is about 0.041 $\mu_B$. This value increases to 0.184 $\mu_B$ at $p = 1.0$. In the latter case, we observe the appearance of spin density on other O sites away from the Ti-CLSs [Fig. 7(h)]. It is notable that the induced magnetic moments on all O atoms are positive, indicating a long-range ferromagnetic type of coupling between the localized moments at the Ti-CLSs and the "continuum" spin density on the other O sites. This indirectly points out the ferromagnetic exchange interaction between the localized magnetic moments mediated by the itinerant electrons. Thus, our results demonstrate the emergence of moiré magnetism in a hole-doped twisted oxide system which is attributed to the O-$p$ states and may have implications for unconventional spin-dependent properties.

Our DFT calculations reveal a tendency for the formation of magnetic moments in the flat bands in twisted $SrTiO_3$ structures. Strong Coulomb correlations within the impurity levels are expected to enhance this trend due to the Anderson mechanism, as long as the localized levels are partially occupied. However, DFT also tends to overestimate the energy of a localized state due to its spurious self-interaction (see Appendix A for more details). Therefore, the states corresponding to the flat band in the AB-stacked $SrTiO_3$ model may in reality lie inside the valence band and hybridize with it. Nevertheless, as seen from our results for the AA-stacked $SrTiO_3$, sufficient hole doping can place the Fermi energy into the deeper flat bands, resulting in exchange splitting and the formation of ferromagnetically-coupled magnetic moments exhibiting moiré periodicity.

## VII. DISCUSSION

We believe that the predicted effects arising in the twisted $SrTiO_3$ bilayers are only the tip of the iceberg of the emergent phenomena which can occur in the moiré-engineered oxide heterostructures. Due to the interplay between the twist angle, local atomic coordination, interface bonding, and electronic and magnetic structure, twisted oxides offer an ample of opportunities for exploring unconventional properties and unknown phenomena.

The observation of vortex polar states at the twisted interface suggests that the system may host topological features. Due to its topological nature, a vortex cannot be continuously deformed into a trivial state without breaking symmetry. This makes vortex structures stable against small perturbations. In contrast to the previously observed polar textures in ferroelectric films confined in the growth direction [54], a polar landscape of twisted oxides is two dimensional and tunable by the twist angle of the bilayer and, thus, it is more amenable for high-density memory applications. In addition, it is known that nanoscale ferroelectrics might exhibit various exotic domain configurations and polar topologies, such as full flux-closure, vortex, skyrmion, and meron [63]. Given the unique structural variations driven by twist, it would be interesting to explore if such topological objects can occur at twisted interfaces. They may also undergo unusual phase transitions and form collective topological polar states under external stimuli.

The emergence of electronic flat bands indicates that twisted $SrTiO_3$ could exhibit strongly correlated properties which deserve further investigation. Superconductivity, in particular, is often associated with flat-band systems, as seen, for example, from the studies of twisted bilayer graphene. [8] As has been predicted theoretically, a flat band coupled with a dispersive band supports high-temperature superconductivity. [64] Furthermore, when the lowest energy band is flat across the entire Brillouin zone, its partial filling is expected to lead to nontrivial behavior, including the formation of a Wigner crystal. [29] In such systems, the vanishing kinetic energy of electrons in the flat band causes their dynamics to be dominated entirely by interactions, driving fermions to self-organize into a regular, crystalline pattern to minimize the interaction energy.

The emergence of moiré magnetism in a hole-doped twisted oxide system may have implications for nontrivial spin-dependent properties. For example, we see that twisted $SrTiO_3$ bilayers produce a sharp peak in the density of states near the valence band maximum. Upon hole doping, the Fermi level can be brought to this elevated density of states, creating favorable conditions for $d^0$ ferromagnetism. It has been recognized that the phenomenon of $d^0$ magnetism in oxides merits thoughtful attention. [65] From the practical perspective, by utilizing ferromagnetism mediated by holes, it may be feasible to control the magnetic state of devices through a gate bias, which modulates the flow of spin-polarized carriers. This method could effectively switch or modify the magnetic state of the system, offering a significant reduction in power consumption compared to conventional spintronic devices that rely on high spin-polarized currents, such as spin-transfer torque. [66] Although hole doping of oxides is generally more challenging than electron doping, the realization of a $p$-type interface in SrTiO$_3$-based heterostructures has been experimentally demonstrated, offering a viable pathway to explore and control hole-mediated ferromagnetism in oxide systems. [67, 68] These findings open new opportunities for the development of energy-efficient, room-temperature spintronic applications.

A logical progression would be to extend these studies to other oxide materials beyond SrTiO$_3$. Especially interesting are oxide compounds exhibiting magnetism in their pristine form. Twisted heterostructures made of magnetic oxides may reveal intricate interface spin textures varying with moiré periodicity. The interlayer exchange coupling is known to be highly sensitive to the oxidation state of magnetic atoms, as well as the bond length and bond angle between the magnetic and oxygen atoms. A lateral modulation of the interlayer exchange coupling is therefore expected to occur at the interface between two twisted magnetic oxides. For example, the interlayer exchange coupling varying between ferromagnetic and antiferromagnetic along the interface tends to generate spin textures, such as isolated skyrmions and skyrmionic lattices, as has been predicted for 2D vdW magnets [69,70].

Furthermore, broken inversion symmetry at the interface between two twisted oxides in conjunction with spin-orbit coupling are expected to produce the Dzyaloshinskii-Moriya interaction (DMI), which promotes nontrivial chiral spin textures, like those predicted for 2D vdW magnets (e.g., [71]). Previous studies have shown that magnetic skyrmions and other spin textures may occur in twisted 2D vdW systems [ 72-75 ]. Similar and more intricate spin structures are expected in twisted oxide bilayers, driven by stronger interlayer interactions. Oxide membranes made of magnetic BiFeO$_3$ and SrRuO$_3$ have already been synthesized down to a monolayer [45, 76] and demonstrated that at reduced dimensions these materials exhibit different properties compared to their bulk counterparts.

Fabrication of twisted transition-metal oxide systems is challenging due to the strong interlayer coupling typically found in oxides. However, recent advances in fabrication techniques, such as the use of sacrificial buffer layers [44,45] and remote epitaxy [46], have established a robust platform for integrating oxide monolayers into twisted heterostructures. The recent experimental work has demonstrated that twisted SrTiO$_3$ bilayers of high structural quality and precisely controlled twist angle and interface atomic structure can be successfully synthesized, showing some of the features predicted in this paper [77]. It is therefore evident that the twisted structures investigated in this work are feasible in practice.

## VIII. SUMMARY

In this study, we have designed SrTiO$_3$ bilayers with a relative twist between the individual layers to explore a broad landscape of emergent phenomena which may occur in moiré-engineered oxide heterostructures. Using extensive DFT calculations, we found that displacement of Ti ions with respect to their unrelaxed positions formed vortex-like patterns. The vortices had circular- or cylindrical-like shapes, exhibited alternating chirality, and were interleaved with antivortices, both forming periodic patterns with moiré periodicity. The formation of the vortex-antivortex displacement patterns indicated the emergence of a chiral polar phase arising from twist and driven by the flexoelectric coupling effect.

Twisting of the SrTiO$_3$ bilayer and associated atomic relaxations led to significant reconstruction of the Ti-O bonds within the TiO$_2$ layers adjacent to the twisted interface. Some of these bonds were found to be broken leading to reduced coordination of the interfacial Ti atoms and, thus, to their reduced oxidation state. The moiré modulation of the Ti oxidation state manifested Ti ionic charge modulation at the twisted interface.

We have further predicted the appearance of flat electronic bands occurring at relatively large twist angles and exhibiting charge localization with moiré periodicity. Analysis of band-decomposed charge densities revealed the localized charge distribution surrounding dangling bonds on the O atoms above the twisted SrTiO$_3$ bilayer interface. Our in-depth tight-binding calculations delved into the origin of these flat bands, revealing that their emergence was facilitated by the broken bonds between O and Ti atoms on a single O site in the moiré supercell.

Finally, we have demonstrated the emergence of unconventional $d^0$ magnetism in response to hole doping of twisted SrTiO$_3$ bilayers. The magnetism was found to be driven by the exchange interactions within the high-density O-$p$ bands resulting in their spin splitting which was reasonably well explained by the Stoner model. The involvement of flat bands in this mechanism led to a localized spin density exhibiting moiré periodicity.

Overall, our results suggest promising avenues for moiré-engineered oxide heterostructures to reveal new properties that are not present in conventional oxide thin-film structures and may be interesting from the fundamental physics point of view and useful for novel applications.

## ACKNOWLEDGEMENTS

This work was supported by the National Science Foundation through EPSCoR RII Track-1 program (NSF Award OIA-2044049) (N.A.S., K.H., E.Y.T), the Office of Naval Research (ONR grant N00014-20-1-2844) (K.S., E.Y.T.), and UNL Grand Challenges catalyst award "Quantum Approaches Addressing Global Threats" (M.E., K.D.B., E.Y.T.). Computations were performed at the University of Nebraska Holland Computing Center. The figures of atomic structures were produced using VESTA software. [78]

## APPENDIX A: DFT CALCULATIONS

Density functional theory (DFT) calculations are performed using the Vienna *ab-initio* simulation package (VASP). [79,80]

We employ the local density approximation (LDA) for the exchange-correlation potential and the projected augmented plane wave (PAW) method to approximate the electron-ion potential. [81] A kinetic energy cutoff of 500 eV for the plane wave expansion of the PAWs is used. Spin-orbit coupling is not included in the calculations.

In the twisted $SrTiO_3$ bilayer calculations, we construct a seven-unit-cell slab terminated with SrO-monolayers having either AA- [Fig. 1(b)] or AB- [Fig. 1(c)] stacking orders along the twisting [001] axes. The upper three monolayers are twisted relative to the lower four monolayers, and the four monolayers adjacent to the twisted interface are relaxed. Additionally, we fix the supercell boundaries, based on the LDA calculated lattice constant of 3.86Å, to prevent atoms from rotating back to their original untwisted positions. We use an 8×8×1 grid of $k$ points for Brillouin zone integration and relax the supercell until forces converged below $10^{-3}$ eV Å$^{-1}$.

To simulate the effects of hole doping, we reduce the number of valence electrons relative to the charge-neutral system. The electron count is adjusted to achieve the desired doping concentration, and charge neutrality is maintained by introducing a uniform compensating background charge. Band occupancies are computed using the tetrahedron method with Blöchl corrections [82], effectively eliminating the broadening effects. The validity of these calculations is verified by using the virtual crystal approximation (VCA) [83] to model the electronic structure of $SrTiO_{3-x}N_x$, replicating the conditions of the hole-doping simulations. The results from both approaches are found to be consistent, confirming the reliability of the doping method for our system.

**Self-interaction effects.** A shift of the flat bands down in energy toward other valence bands with increasing hole doping is due to the self-interaction effect known from standard density functional theory (see, e.g., [84] for review). This shift arises because an electron interacts with its own charge density, leading to a somewhat incorrect potential. Flat band systems are more prone to the self-interaction effect because of the high charge density localized in space. Specifically, the flat band shifts lower in energy with the increased hole doping due to the proportionally reduced number of electrons occupied by this band and thus a reduced self-interaction energy.

## APPENDIX B: POLARIZATION VORTICES IN A TWISTED $SrTiO_3$ BILAYER WITH AB STACKING

Similar to the AA-stacked bilayer, a twisted $SrTiO_3$ bilayer with AB stacking reveals vortex-antivortex patterns in displacement of Ti atoms [Figs. A1(a,e)]. The vortices and antivortices exhibit alternating chirality, form a periodic pattern with moiré periodicity, and have opposite chirality in the interfacial top and bottom $TiO_2$ monolayers. They produce intralayer shear strain $\varepsilon_{xy}$ with alternating regions of positive and negative $\varepsilon_{xy}$ [Figs. A1(b,f)]. X- and y-components of Ti ion displacements [light gray arrows in Figs. A1(c,d) and Figs. A1(g,h)] reflect $P_x$ and $P_y$ components of electric polarization and correlate with the shear gradients $\partial\varepsilon_{xy}/\partial y$ [Figs. A1(d,h) and $\partial\varepsilon_{xy}/\partial x$ [Figs.

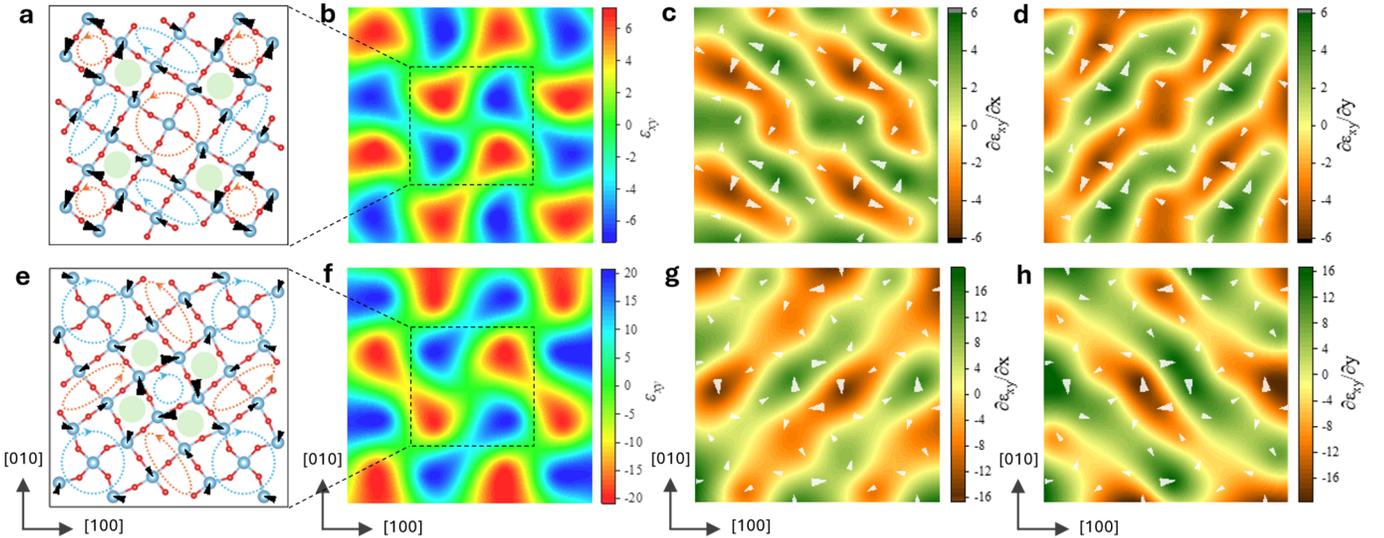

**Fig. A1** Formation of polarization vortices in a 22.6° twisted $SrTiO_3$ bilayer with AB stacking. (a, e) Atomic structure of the $TiO_2$ monolayer nearest to the interface in the top (a) and bottom (e) $SrTiO_3$ layers. Black arrows (proportional in size) show displacement of the Ti atoms with respect to their unrelaxed positions. Dashed circles and ellipses (differed in color and arrow direction) indicate the formation of displacement vortices of opposite chirality. Regions highlighted in green signify antivortices. (b-c, f-h) Shear strain (b, f) and shear strain gradient (c-d, g-f) maps obtained for $TiO_2$ monolayers near the interface in the top (b,c,d) and bottom (f,g,h) $SrTiO_3$ layers. Gray arrows (proportional in size) show displacement of the Ti atoms with respect to their unrelaxed positions.

A1(c,g)], respectively, indicating flexoelectric coupling between the strain gradient and polarization.

## APPENDIX C: DDEC AND BADER CHARGE ANALYSES

Density-Derived Electrostatic and Chemical (DDEC) charges are calculated using the code developed by Manz and co-workers [56, 85-88]. Within this code, the DDEC net atomic charges are optimized to obtain chemical states and quantify electron transfer between atoms in complex materials. Specifically, the DDEC method incorporates spherical averaging to minimize atomic multipole magnitudes so that the electrostatic potential is accurately reproduced by the net atomic charges. The DDEC method uses reference ion densities to provide chemical significance of the computed net atomic charges and includes the effects of charge compensation and dielectric screening in the reference ion densities. The code is parallelized using OpenMP and executable with the charge density output of the VASP calculations.

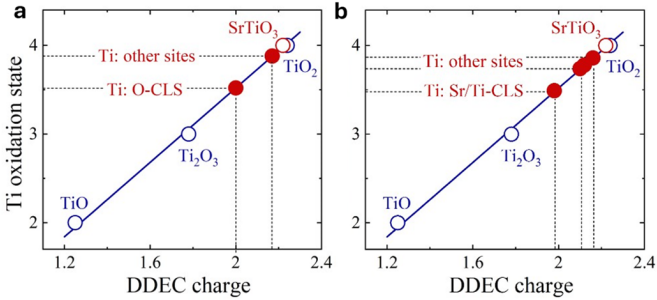

**Fig. A2** Correlation of the nominal Ti oxidation states in 22.6° twisted AA- (a) and AB- (b) stacked $SrTiO_3$ bilayers with the calculated DDEC charges on Ti atoms in bulk TiO, $Ti_2O_3$, and $TiO_2$ compounds where the nominal Ti oxidation states are 2+, 3+, and 4+, respectively (open blue dots), and in bulk $SrTiO_3$ where the Ti oxidation state is 4+ (open red dot). Solid line is a linear fit from which the oxidation state of the interfacial Ti atoms in the twisted $SrTiO_3$ bilayers is estimated (solid red dots).

Figure A2(a) shows the calculated DDEC charges of bulk TiO, $Ti_2O_3$, and $TiO_2$ (open blue circles), which we correlate with the conventional oxidation states of these compounds (2+, 3+, and 4+, respectively). In addition, in Figure A2(a), we display the DDEC charge of bulk $SrTiO_3$ (open red circle) which has the same nominal oxidation state 4+ as $TiO_2$ and nearly the same DDEC charge. Using these results, we obtain a calibration curve [straight line in Figure A2(a)], correlating the calculated DDEC charges with the Ti oxidation states. Then, using this calibration curve along with the computed DDEC charges in twisted $SrTiO_3$ bilayers, we estimate the oxidation state of the interfacial Ti atoms. For a 22.6° twisted AA-stacked $SrTiO_3$ bilayer, we find the oxidation state of 3.51+ for the 5-fold coordinated Ti atoms located at the O-CLSs and 3.89+ for 6-fold coordinated Ti atoms including those located at the Ti- or Sr-CLSs [Fig. A2(a), solid red dots]. For an AB-stacked $SrTiO_3$ bilayer [Fig. A2(b)], we find the oxidation state of 3.51+ for the 5-fold coordinated Ti atoms located at the Sr/Ti-CLSs and 3.86+ for 6-fold coordinated Ti atoms located at the Ti/Sr-CLSs. For other 6-fold coordinated Ti atoms the oxidation states are 3.74+ and 3.78+ [Fig. A2(b), solid red dots].

Bader charges are calculated using the code of Tang *et al.* [89]. These calculations are based on a computational method that partitions the electron density into atomic regions where the dividing surfaces are at a minimum of the charge density, i.e. the gradient of the charge density is zero along the surface normal. The total charge on an atom is obtained by integrating the electronic density within these partitions around the atomic nuclei. The Bader charge for an atom is then defined as the difference between the number of its valence electrons and its total charge obtained from the Bader charge analysis [57]. Correlating the calculated Bader charges in the twisted $SrTiO_3$ bilayers with the known nominal oxidation states of Ti atoms and the calculated Bader charges in the reference oxide compounds, we find a very similar behavior to that obtained from the DDEC charge analysis. We therefore do not explicitly reproduce these results here.

## APPENDIX D: TOPMOST VALENCE BAND IN AA- STACKED TWISTED $SrTiO_3$ BILAYER

Figure A3(a) shows the electronic band structure of a 22.6° twisted AA-stacked $SrTiO_3$ bilayer in the vicinity of the valence band maximum. The topmost valence band is dispersive, and its charge density is uniformly distributed in the plane of the bilayer (Fig. A3(b), bottom panel). It is evident from the top panel in Fig. A3(b) that this band is largely localized in the interfacial TiO$_2$ monolayer. Figure A3(c) demonstrates that its DOS (corresponding to the energy window down to −0.1 eV) is relatively low (about 10 eV$^{-1}$ corresponding to about 0.36 eV$^{-1}$ per O atom), reflecting its free-electron-like nature.

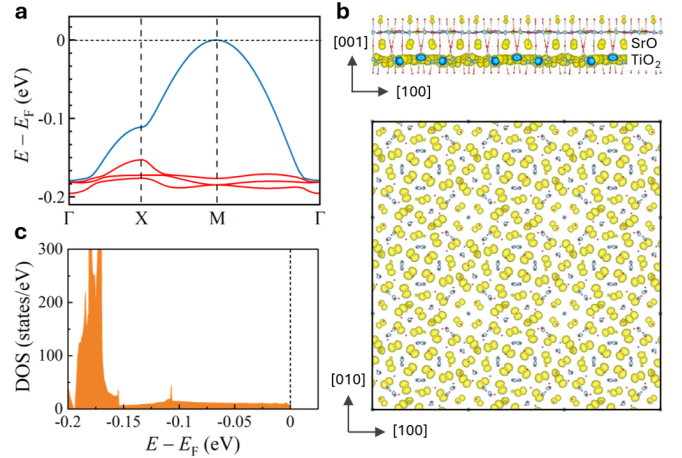

**Fig. A3** Electronic properties of the dispersive band in a 22.6° twisted AA-stacked $SrTiO_3$ bilayer. (a) Electronic band structure along high symmetry lines in the first Brillouin zone at the topmost valence band. Blue and red curves indicate the dispersive and flat bands, respectively. (b) Isosurface of the charge density (shown in yellow) in real space for the dispersive band. Top: side view; bottom: top view. Interfacial SrO and $TiO_2$ monolayers are indicated. (c) Total density of states (DOS) at energies close to the VBM. Dashed line indicates the position of the Fermi energy.

## APPENDIX E: DOS OF AA-STACKED TWISTED SrTiO$_3$ BILAYER AS A FUNCTION OF DOPING

Figure A4 shows the evolution of DOS of a non-spin-polarized AA-stacked twisted SrTiO$_3$ bilayer with hole doping. Filling the topmost valence band with holes shifts the Fermi energy down in energy. At the same time, we observe the displacement of the flat bands to higher energies closer to the VBM. At $p = 0.7$, hope doping places the Fermi energy at about 0.045 eV below the VBM at the top edge of the flat bands where the DOS of states is enhanced up to about 75 eV$^{-1}$ [inset in Fig. A4(d)] which is sufficient to satisfy the Stoner criterion.

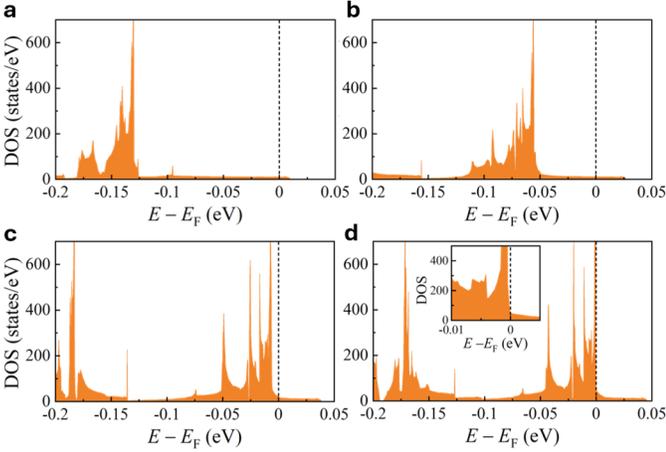

**Fig. A4** Total density of states (DOS) of a non-spin-polarized twisted SrTiO$_3$ bilayer with AA stacking at energies close to the VBM for $p = 0.1$ (a), 0.3 (b), 0.5 (c), and 0.7 (d). Vertical dashed lines indicate the position of the Fermi energy. Inset in (d) shows a zoomed-in region near the Fermi energy.

## APPENDIX F: TIGHT-BINDING MODELING

We employ a standard TB model that assumes one orbital per atom and nearest-neighbor hopping of a periodic lattice. The TB Hamiltonian in real space is given by

$$H = \sum_{m\alpha} \varepsilon_\alpha c^\dagger_{m\alpha} c_{m\alpha} - \sum_{\langle m\alpha, l\beta \rangle} t_{m\alpha, l\beta} c^\dagger_{m\alpha} c_{l\beta}, \qquad (1)$$

where $c^\dagger_{m\alpha}$ and $c_{m\alpha}$ are the creation and annihilation operators at cell $p$ and atomic site $\alpha$, $\varepsilon_\alpha$ are on-site atomic energies, and $t_{m\alpha,l\beta}$ hopping integrals between sites $\alpha$ and $\beta$ in cells $m$ and $l$, respectively, and summation $\langle m\alpha, l\beta \rangle$ runs over the nearest-neighbor sites. The eigenstates $|\psi_{n\mathbf{k}}\rangle$ are represented through an expansion over atomic orbitals $|\varphi_{m\alpha}\rangle$:

$$|\psi_{n\mathbf{k}}\rangle = \sum_{m\alpha} e^{i\mathbf{k}\cdot(\mathbf{R}_m+\mathbf{r}_\alpha)} C^\alpha_{n\mathbf{k}} |\varphi_{m\alpha}\rangle, \qquad (2)$$

where $n$ is the band index, $\mathbf{k}$ is the wave vector, and $\mathbf{R}_m$ is the coordinate on the lattice cell $m$, $\mathbf{r}_\alpha$ is the coordinate of atom $\alpha$ in the unit cell of the lattice. The expansion coefficients $C^\alpha_{n\mathbf{k}}$ are obtained by solving the TB secular equation

$$\sum_\beta H_{\alpha\beta}(\mathbf{k}) C^\beta_{n\mathbf{k}} = \varepsilon_{n\mathbf{k}} C^\alpha_{n\mathbf{k}}, \qquad (3)$$

where the matrix elements of the TB Hamiltonian in $k$-space are

$$H_{\alpha\beta}(\mathbf{k}) = \sum_m e^{i\mathbf{k}\cdot(\mathbf{R}_m+\mathbf{r}_\beta-\mathbf{r}_\alpha)} \langle \varphi_{0\alpha} | H | \varphi_{m\beta} \rangle. \qquad (4)$$

**1D TB model.** Green's function formalism is used for a one-dimensional (1D) TB model depicted in Figure 3(a) to understand the origin of flat bands. We start from an infinite double wire that has two atomic sites ($\alpha, \beta = 1,2$) with spacing $a$ between the sites. The sites alternate between each lattice cells and have on-site energies $\varepsilon_\alpha = \mp\Delta$ [Fig. A5(a)]. The wire is described by TB Hamiltonian (1), where we assume that hopping integrals $t = t_{m\alpha,l\beta}$ between nearest neighbors are equal to 1, i.e. all energies are "measured" in units of this hopping parameter. Site alternation can be incorporated into the eigen-state $|\psi_{n\mathbf{k}}\rangle$ by introducing the lattice site switching operator $\sigma$, which flips site index $\alpha$ ($\alpha = 1,2$), without affecting lattice index $m$:

$$|\psi_{n\mathbf{k}}\rangle = \sum_{m\alpha} e^{ikR_m} \sigma^m C^\alpha_{n\mathbf{k}} |\varphi_{m\alpha}\rangle, \qquad (5)$$

where $R_m = am$. Taking into account that $\sigma^{2m} = 1$ and $\sigma^{2m+1} = \sigma$, we obtain the following matrix elements of the TB Hamiltonian:

$$H_{\alpha\beta}(k) = \begin{bmatrix} -\Delta & -2\cos ka - 1 \\ -2\cos ka - 1 & \Delta \end{bmatrix}. \qquad (6)$$

By solving secular equations (3), we obtain eigen-values $\varepsilon_{nk}$ and coefficients $C^\alpha_{nk}$ in expansion (5) as follows:

$$\varepsilon_{nk} = \pm\sqrt{\Delta^2 + (2\cos ka + 1)^2}, \qquad (7)$$

$$C^1_{nk} = -\sqrt{\frac{\varepsilon_{nk} - \Delta}{2\varepsilon_{nk}}}, \quad C^2_{nk} = \sqrt{\frac{\varepsilon_{nk} + \Delta}{2\varepsilon_{nk}}}, \qquad (8)$$

where $n = 1,2$.

Real-space components of Green's function are given by

$$G_{m\alpha,l\beta}(z) = \left\langle \varphi_{m\alpha} \left| \frac{1}{z-H} \right| \varphi_{m\beta} \right\rangle = \sum_{nk} \frac{e^{ika(m-l)}}{z - \varepsilon_{nk}} C^\alpha_{nk} C^{\beta*}_{nk}. \qquad (9)$$

By performing integration over $k$ and using Eqs. (7) and (8), we obtain Green's function components within the diatomic unit cell ($l = m$):

$$G_{\alpha\alpha}(E) = \frac{E \pm \Delta}{\sqrt{(E^2 - \Delta^2)}} \left[ \frac{\chi_1(E)}{\left(\sqrt{E^2 - \Delta^2} + 1\right)^{1/2} \left(\sqrt{E^2 - \Delta^2} - 3\right)^{1/2}} \right.$$
$$\left. + \frac{\chi_2(E)(\sqrt{E^2 - \Delta^2} + 1)}{4\left(\sqrt{E^2 - \Delta^2} - 1\right)^{1/2} \left(\sqrt{E^2 - \Delta^2} + 3\right)^{1/2}} \right], \quad (10)$$

where sign is positive (negative) for $\alpha = 1$ ($\alpha = 2$), and

$$G_{\alpha\beta}(E) = -\frac{1}{2} + \frac{\chi_1(E)\left(\sqrt{E^2 - \Delta^2} - 1\right)}{4\left(\sqrt{E^2 - \Delta^2} + 1\right)^{1/2}\left(\sqrt{E^2 - \Delta^2} - 3\right)^{1/2}}$$
$$+ \frac{\chi_2(E)(\sqrt{E^2 - \Delta^2} + 1)}{4\left(\sqrt{E^2 - \Delta^2} - 1\right)^{1/2}\left(\sqrt{E^2 - \Delta^2} + 3\right)^{1/2}}, \quad (11)$$

where $\alpha \neq \beta$. In Eqs. (10) and (11),

$$\chi_\alpha(E) = \begin{cases} 1, & \text{for } |E| > \sqrt{\Delta^2 + \gamma^2}, \\ sgn(E), & \text{for } |E| < \sqrt{\Delta^2 + \gamma^2}, \end{cases} \quad (12)$$

where $\gamma = 3$ for $\alpha = 1$ and $\gamma = 1$ for $\alpha = 2$. In the interval of energies $\Delta \leq |E| \leq \sqrt{\Delta^2 + 9}$, Eq. (10) results in a non-zero site-projected density of states $\rho_\alpha(E) = -\frac{1}{\pi} \text{Im } G_{\alpha\alpha}(E)$, while beyond this interval $\rho_\alpha(E) = 0$. For non-zero $\Delta$, $\rho_\alpha(E)$ is continuous in the regions $-\sqrt{\Delta^2 + 9} \leq E \leq -\Delta$ and $\Delta \leq E \leq \sqrt{\Delta^2 + 9}$, representing two bands separated by a band gap of $2\Delta$.

Next, we construct surface Green's function $G^s$ of a semi-infinite double wire. This is done by connecting two such wires with the same termination to a diatomic cell with hopping $t$ linking sites 1 to 2, as shown in Figure A5(b). This procedure reconstructs the infinite double wire [Fig. A5(a)], whose Green's function $G$ is given by Eqs. (10-13). Using the Dyson equation, we obtain

$$G = [G^{c-1} - 2VG^sV + i\delta]^{-1}. \quad (13)$$

Here $V = \begin{bmatrix} 0 & -1 \\ -1 & 0 \end{bmatrix}$ (in units of $t$) and $G^c$ is Green's function of a diatomic cell between the chains:

$$G^c_{\alpha\alpha}(E) = \frac{E \mp \Delta}{E^2 - \Delta^2 - 1}, \quad (14)$$

where sign is negative (positive) for $\alpha = 1$ ($\alpha = 2$), and

$$G^c_{\alpha\beta}(E) = \frac{-1}{E^2 - \Delta^2 - 1} \quad (15)$$

for $\alpha \neq \beta$. Solving Eq. (13) for $G^s$, we find

$$G^s = \tfrac{1}{2}\, V^{-1}(G^{c-1} - G^{-1})V^{-1}. \quad (16)$$

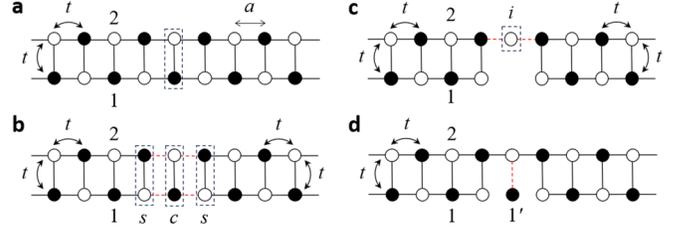

Fig. A5 Schematics of constructing Green's functions within a 1D tight-binding model. (a) Two coupled infinite atomic chains. Dashed box indicates a diatomic unit cell where Green's function is calculated. (b) Two semi-infinite atomic chains coupled through a diatomic cell. Indices $s$ and $c$ indicate surface cells of the semi-infinite chains and a diatomic cell, respectively. (c) Two semi-infinite chains coupled through site 1 indicated by index $i$ denoting the on-site Green's function. (d) Final atomic configuration with atom $1'$ having two broken bonds in the bottom chain. Red dashed lines indicate bonds of strength $t$ that are used to derive Green's functions from the respective Dyson equations.

Next, we construct the on-site Green's function $G^i(E)$ on atom 2 in the atomic structure shown in Figure A5(c). This structure corresponds to the configuration of Figure 5(a) where atom 2 is decoupled from the atom with the dangling bond. The Green's function is

$$G^i(E) = [E - \Delta - 2G^s_{11}(E) + i\delta]^{-1}. \quad (17)$$

Finally, adding bond $t$ to atom $1'$, as shown in Figure A5(d), and using the Dyson equation, we obtain the on-site Green's function of the atom $1'$ with broken in-plane bonds as follows

$$G^{1'}(E) = [E - \varepsilon_{1'} - G^i(E) + i\delta]^{-1}. \quad (18)$$

The on-site energy $\varepsilon_{1'}$ of atom $1'$ is expected to be close to $-\Delta$ reflecting the fact that atom $1'$ is essentially atom 1 placed in an altered atomic environment. Function (18) has singularities corresponding to localized (resonant) states whose energy $E_{loc}$ depends on $\varepsilon_{1'}$. Expanding $G^i(E)$ around $E = -\Delta$, for $\varepsilon_{1'} < -\Delta$, yields

$$E_{loc} \approx -\Delta \left[ 1 \pm \frac{96}{49}(\varepsilon_{1'} + \Delta)^2 \right]. \quad (19)$$

From Eq. (19), we see that for $\varepsilon_{1'} = -\Delta$, the localized state sits on top of the bottom band at $E_{loc} = -\Delta$. For $\varepsilon_{1'} < -\Delta$, but close to it, there are two singular states. One is a localized state in the band gap, and the other is a resonant state inside the continuum. These states are seen in Figure 5(b) as a red line (localized state) and as a peak in DOS just below $E = -0.5$ (resonant state).

**2D TB model.** We consider the 2D atomic structure described in Section V and shown in Figure 5(c). TB calculations are performed using the PythTB code [90]. In the calculations, we consider an 8×8 supercell and TB parameters $t = 1$ and $\Delta = 0.5$.

We find that for $\varepsilon_{1'} = \varepsilon_1 = -0.5$, there is a localized state which is characterized by a quasi-flat band sitting on top of the valence band. When $|\varepsilon_{1'}|$ is reduced, the band moves into the band gap and becomes completely decoupled from the continuum, as shown in Figure 5(d) for $\varepsilon_{1'} = -0.2$. When $|\varepsilon_{1'}|$ is increased so that $|\varepsilon_{1'}| > 0.5$, the state moves down in energy into the valence band continuum and becomes strongly hybridized with the continuum states.

1. J. M. B. Lopes dos Santos, N. M. R. Peres, and A. H. Castro Neto, Graphene bilayer with a twist: Electronic structure. *Phys. Rev. Lett.* **99**, 256802 (2007).
2. R. Bistritzer and A. H. MacDonald, Moiré bands in twisted double-layer graphene. *Proc. Nat. Acad. Sci.* **108**, 12233–12237 (2011).
3. S. Carr, D. Massatt, S. Fang, P. Cazeaux, M. Luskin, and E. Kaxiras, Twistronics: Manipulating the electronic properties of two-dimensional layered structures through their twist angle. *Phys. Rev. B* **95**, 075420 (2017).
4. R. Ribeiro-Palau, C. Zhang, K. Watanabe, T. Taniguchi, J. Hone, and C. R. Dean, Twistable electronics with dynamically rotatable heterostructures. *Science* **361**, 690–693 (2018).
5. D. M. Kennes, M. Claassen, L. Xian, A. Georges, A. J. Millis, J. Hone, C. R. Dean, D. N. Basov, A. N. Pasupathy, and A. Rubio. Moiré heterostructures as a condensed-matter quantum simulator. *Nat. Phys.* **17**, 155–163 (2021).
6. Z. Hennighausen and S. Kar, Twistronics: A turning point in 2D quantum materials, *Electron. Struct.* **3**, 014004 (2021).
7. S. W. Li, K. Wei, Q. R. Liu, Y. X. Tang, and T. Jiang, Twistronics and moiré excitonic physics in van der Waals heterostructures *Front. Phys.* **19**, 42501 (2024).
8. Y. Cao, V. Fatemi, S. Fang, K. Watanabe, T. Taniguchi, E. Kaxiras, and P. Jarillo-Herrero. Unconventional superconductivity in magic-angle graphene superlattices. *Nature* **556**, 43–50 (2018).
9. Z. Zhang, Y. Wang, K. Watanabe, T. Taniguchi, K. Ueno, E. Tutuc, and B. J. LeRoy, Flat bands in twisted bilayer transition metal dichalcogenides. *Nat. Phys.* **16**, 1093–1096 (2020).
10. L. Wang, E.-M. Shih, A. Ghiotto, L. Xian, D. A. Rhodes, C. Tan, M. Claassen, D. M. Kennes, Y. Bai, B. Kim, K. Watanabe, T. Taniguchi, X. Zhu, J. Hone, A. Rubio, A. N. Pasupathy, and C. R. Dean, Correlated electronic phases in twisted bilayer transition metal dichalcogenides. *Nat. Mater.* **19**, 861–866 (2020).
11. M. H. Naik and M. Jain, Ultraflatbands and shear solitons in moiré patterns of twisted bilayer transition metal dichalcogenides. *Phys. Rev. Lett.* **121**, 266401 (2018).
12. E. Li, J.-X. Hu, X. Feng, Z. Zhou, L. An, K. T. Law, N. Wang, and N. Lin, Lattice reconstruction induced multiple ultra-flat bands in twisted bilayer WSe$_2$. *Nat. Commun.* **12**, 5601 (2021).
13. L. Xian, D. M. Kennes, N. Tancogne-Dejean, M. Altarelli, and A. Rubio, Multiflat bands and strong correlations in twisted bilayer boron nitride: Doping-Induced correlated insulator and superconductor. *Nano Lett.* **19**, 4934–4940 (2019).
14. N. R. Walet and F. Guinea, Flat bands, strains, and charge distribution in twisted bilayer $h$-BN. *Phys. Rev. B* **103**, 125427 (2021).
15. D. Kang, Z. Zuo, W. Ju, and Z. Wang, Emergence of flat bands in twisted bilayer C$_3$N induced by simple localization and destructive interference. *Phys. Rev. B* **107**, 85425 (2023).
16. C. F. Li, W. J. Zhai, Y. Q. Li, Y. S. Tang, J. H. Zhang, P. Z. Chen, G. Z. Zhou, X. M. Cui, L. Lin, and Z. B. Yan, Extremely flat band in antiferroelectric bilayer α-In$_2$Se$_3$ with large twist-angle. *New J. Phys.* **23**, 083019 (2021).
17. D. M. Kennes, L. Xian, M. Claassen, and A. Rubio, One-dimensional flat bands in twisted bilayer germanium selenide. *Nat. Commun.* **11**, 1124 (2020).
18. T. Kariyado, Twisted bilayer BC$_3$: Valley interlocked anisotropic flat bands. *Phys. Rev. B* **107**, 85127 (2023).
19. H. Li, S. Li, M. H. Naik, J. Xie, X. Li, J. Wang, E. Regan, D. Wang, W. Zhao, S. Zhao, S. Kahn, K. Yumigeta, M. Blei, T. Taniguchi, K. Watanabe, S. Tongay, A. Zettl, S. G. Louie, F. Wang, and M. F. Crommie, Imaging moiré flat bands in three-dimensional reconstructed WSe$_2$/WS$_2$ superlattices. *Nat. Mater.* **20**, 945–950 (2021).
20. B. L. Chittari, G. Chen, Y. Zhang, F. Wang, and J. Jung, Gate-tunable topological flat bands in trilayer graphene boron-nitride moiré superlattices. *Phys. Rev. Lett.* **122**, 16401 (2019).
21. S. Xie, B. D. Faeth, Y. Tang, L. Li, E. Gerber, C. T. Parzyck, D. Chowdhury, Y.-H. Zhang, C. Jozwiak, A. Bostwick, E. Rotenberg, E.-A. Kim, J. Shan, K. F. Mak, and K. M. Shen, Strong interlayer interactions in bilayer and trilayer moiré superlattices. *Sci. Adv.* **8**, eabk1911 (2023).
22. J. Yang, G. Chen, T. Han, Q. Zhang, Y.-H. Zhang, L. Jiang, B. Lyu, H. Li, K. Watanabe, T. Taniguchi, Z. Shi, T. Senthil, Y. Zhang, F. Wang, and L. Ju, Spectroscopy signatures of electron correlations in a trilayer graphene/hBN moiré superlattice. *Science* **375**, 1295–1299 (2022).
23. Y. Li, S. Zhang, F. Chen, L. Wei, Z. Zhang, H. Xiao, H. Gao, M. Chen, S. Liang, D. Pei, L. Xu, K. Watanabe, T. Taniguchi, L. Yang, F. Miao, J. Liu, B. Cheng, M. Wang, Y. Chen, and Z. Liu. Observation of coexisting Dirac bands and Moiré flat bands in magic-angle twisted trilayer graphene. *Adv. Mater.* **34**, 2205996 (2022).
24. G. Chen, A. L. Sharpe, P. Gallagher, I. T. Rosen, E. J. Fox, L. Jiang, B. Lyu, H. Li, K. Watanabe, T. Taniguchi, J. Jung, Z. Shi, D. Goldhaber-Gordon, Y. Zhang, and F. Wang, Signatures of tunable superconductivity in a trilayer graphene moiré superlattice. *Nature* **572**, 215–219 (2019).
25. C. Repellin, Z. Dong, Y.-H. Zhang, and T. Senthil, Ferromagnetism in narrow bands of moiré superlattices. *Phys. Rev. Lett.* **124**, 187601 (2020).
26. G. Chen, A. L. Sharpe, E. J. Fox, Y.-H. Zhang, S. Wang, L. Jiang, B. Lyu, H. Li, K. Watanabe, T. Taniguchi, Z. Shi, T. Senthil, D. Goldhaber-Gordon, Y. Zhang, and F. Wang, Tunable correlated Chern insulator and ferromagnetism in a moiré superlattice. *Nature* **579**, 56–61 (2020).
27. J.-X. Lin, Y.-H. Zhang, E. Morissette, Z. Wang, S. Liu, D. Rhodes, K. Watanabe, T. Taniguchi, J. Hone, and J. I. A. Li, Spin–orbit-driven ferromagnetism at half moiré filling in magic-angle twisted bilayer graphene. *Science* **375**, 437–441 (2022).

28. G. Chen, A. L. Sharpe, E. J. Fox, S. Wang, B. Lyu, L. Jiang, H. Li, K. Watanabe, T. Taniguchi, M. F. Crommie, M. A. Kastner, Z. Shi, D. Goldhaber-Gordon, Y. Zhang, and F. Wang, Tunable orbital ferromagnetism at noninteger filling of a moiré superlattice. *Nano Lett.* **22**, 238–245 (2022).
29. C. Wu, D. Bergman, L. Balents, and S. Das Sarma, Flat bands and wigner crystallization in the honeycomb optical lattice. *Phys. Rev. Lett.* **99**, 70401 (2007).
30. M. Serlin, C. L. Tschirhart, H. Polshyn, Y. Zhang, J. Zhu, K. Watanabe, T. Taniguchi, L. Balents, and A. F. Young, Intrinsic quantized anomalous Hall effect in a moiré heterostructure. *Science* **367**, 900–903 (2020).
31. N. Stefanidis and I. Sodemann, Excitonic Laughlin states in ideal topological insulator flat bands and their possible presence in moiré superlattice materials. *Phys. Rev. B* **102**, 35158 (2020).
32. B. Liu, L. Xian, H. Mu, G. Zhao, Z. Liu, A. Rubio, and Z. F. Wang Higher-order band topology in twisted moiré superlattice. *Phys. Rev. Lett.* **126**, 66401 (2021).
33. H. Polshyn, Y. Zhang, M. A. Kumar, T. Soejima, P. Ledwith, K. Watanabe, T. Taniguchi, A. Vishwanath, M. P. Zaletel, and A. F. Young, Topological charge density waves at half-integer filling of a moiré superlattice. *Nat. Phys.* **18**, 42–47 (2022).
34. Z. Zheng, Q. Ma, Z. Bi, S. de la Barrera, M.-H. Liu, N. Mao, Y. Zhang, N. Kiper, K. Watanabe, T. Taniguchi, J. Kong, W. A. Tisdale, R. Ashoori, N. Gedik, L. Fu, S.-Y. Xu, and P. Jarillo-Herrero, Unconventional ferroelectricity in moiré heterostructures. *Nature* **588**, 71–76 (2020).
35. D. Chen, Z. Lian, X. Huang, Y. Su, M. Rashetnia, L. Ma, L. Yan, M. Blei, L. Xiang, T. Taniguchi, K. Watanabe, S. Tongay, D. Smirnov, Z. Wang, C. Zhang, Y.-T. Cui, and S.-F. Shi, Excitonic insulator in a heterojunction moiré superlattice, *Nat. Phys.* **18**, 1171–1176 (2022).
36. L. Du, M. R. Molas, Z. Huang, G. Zhang, F. Wang, and Z. Sun, Moiré photonics and optoelectronics. *Science* **379**, eadg0014 (2023).
37. L. Du, Z. Huang, J. Zhang, F. Ye, Q. Dai, H. Deng, G. Zhang, and Z. Sun, Nonlinear physics of moiré superlattices. *Nat. Mater.* **23**, 1179–1192 (2024).
38. C.-C. Lee, A. Fleurence, Y. Yamada-Takamura, and T. Ozaki, Hidden mechanism for embedding the flat bands of Lieb, kagome, and checkerboard lattices in other structures. *Phys. Rev. B* **100**, 45150 (2019).
39. L. Zheng, L. Feng, and W. Yong-Shi, Exotic electronic states in the world of flat bands: From theory to material. *Chinese Phys. B* **23**, 77308 (2014).
40. M. Kang, S. Fang, L. Ye, H. C. Po, J. Denlinger, C. Jozwiak, A. Bostwick, E. Rotenberg, E. Kaxiras, J. G. Checkelsky, and R. Comin, Topological flat bands in frustrated Kagome lattice CoSn. *Nat. Commun.* **11**, 4004 (2020).
41. A. Julku, S. Peotta, T. I. Vanhala, D.-H. Kim, and P. Törmä, Geometric origin of superfluidity in the Lieb-lattice flat band. *Phys. Rev. Lett.* **117**, 45303 (2016).
42. S. Mukherjee, A. Spracklen, D. Choudhury, N. Goldman, P. Öhberg, E. Andersson, and R. R. Thomson, Observation of a localized flat-band state in a photonic Lieb lattice. *Phys. Rev. Lett.* **114**, 245504 (2015).
43. M. Iskin, Origin of flat-band superfluidity on the Mielke checkerboard lattice. *Phys. Rev. A* **99**, 53608 (2019).
44. D. Lu, D. J. Baek, S. S. Hong, L. F. Kourkoutis, Y. Hikita, and H. Y. Hwang, Synthesis of freestanding single-crystal perovskite films and heterostructures by etching of sacrificial water-soluble layers. *Nat. Mater.* **15**, 1255–1260 (2016).
45. D. Ji, S. Cai, T. R. Paudel, H. Sun, C. Zhang, L. Han, Y. Wei, Y. Zang, M. Gu, Y. Zhang, W. Gao, H. Huyan, W. Guo, D. Wu, Z. Gu, E. Y. Tsymbal, P. Wang, Y. Nie, and X. Pan, Freestanding crystalline oxide perovskites down to the monolayer limit. *Nature* **570**, 87–90 (2019).
46. H. S. Kum, H. Lee, S. Kim, S. Lindemann, W. Kong, K. Qiao, P. Chen, J. Irwin, J. H. Lee, S. Xie, S. Subramanian, J. Shim, S.-H. Bae, C. Choi, L. Ranno, S. Seo, S. Lee, J. Bauer, H. Li, K. Lee, J. A. Robinson, C. A. Ross, D. G. Schlom, M. S. Rzchowski, C.-B. Eom, and J. Kim, Heterogeneous integration of single-crystalline complex-oxide membranes. *Nature* **578**, 75–81 (2020).
47. Y. Li, C. Xiang, F. M. Chiabrera, S. Yun, H. Zhang, D. J. Kelly, R. T. Dahm, C. K. R. Kirchert, T. E. Le Cozannet, F. Trier, D. V. Christensen, T. J. Booth, S. B. Simonsen, S. Kadkhodazadeh, T. S. Jespersen, and N. Pryds, Stacking and twisting of freestanding complex oxide thin films. *Adv. Mater.* **34**, 2203187 (2022).
48. J. Shen, Z. Dong, M. Q. Qi, Y. Zhang, C. Zhu, Z. Wu, and D. Li, Observation of moiré patterns in twisted stacks of bilayer perovskite oxide nanomembranes with various lattice symmetries. *ACS Appl. Mater. Interfaces* **14**, 50386–50392 (2022).
49. G. Sánchez-Santolino, V. Rouco, S. Puebla, H. Aramberri, V. Zamora, M. Cabero, F. A. Cuellar, C. Munuera, F. Mompean, M. Garcia-Hernandez, A. Castellanos-Gomez, J. Íñiguez, C. Leon, and J. Santamaria, A 2D ferroelectric vortex pattern in twisted $BaTiO_3$ freestanding layers. *Nature* **626**, 529–534 (2024).
50. V. Vitale, K. Atalar, A. A. Mostofi, and J. Lischner, Flat band properties of twisted transition metal dichalcogenide homo- and heterobilayers of $MoS_2$, $MoSe_2$, $WS_2$ and $WSe_2$. *2D Mater.* **8**, 045010 (2021).
51. N. Lu, H. Guo, Z. Zhuo, L. Wang, X. Wu, and X. C. Zeng, Twisted $MX_2/MoS_2$ heterobilayers: Effect of van der Waals interaction on the electronic structure. *Nanoscale* **9**, 19131–19138 (2017).
52. Q. Xu, Y. Guo, and L. Xian, Moiré flat bands in twisted 2D hexagonal vdW materials. *2D Mater.* **9**, 014005 (2022).
53. Z. Song, Y. Wang, H. Zheng, P. Narang, and L.-W. Wang, Deep quantum-dot arrays in moiré superlattices of non-van der Waals materials. *J. Am. Chem. Soc.* **144**, 14657–14667 (2022).
54. A. K. Yadav, C. T. Nelson, S. L. Hsu, Z. Hong, J. D. Clarkson, C. M. Schlepütz, A. R. Damodaran, P. Shafer, E. Arenholz, L. R. Dedon, D. Chen, A. Vishwanath, A. M. Minor, L. Q. Chen, J. F. Scott, L. W. Martin, and R. Ramesh, Observation of polar vortices in oxide superlattices. *Nature* **530**, 198–201 (2016).
55. P. Zubko, G. Catalan, and A. K. Tagantsev, Flexoelectric effect in solids. *Annu. Rev. Mater. Res.* **43**, 387–421 (2013).
56. T. A. Manz and D. S. Sholl, Chemically meaningful atomic charges that reproduce the electrostatic potential in periodic and nonperiodic materials. *J. Chem. Theory Comput.* **6**, 2455–2468. (2010).
57. R. F. W. Bader, *Atoms in Molecules: A Quantum Theory* (Oxford Univ. Press, New York, 1994).
58. Y. A. Izyumov, Theory of an impurity state in a crystal. *Adv. Phys.* **14**, 569–619 (1965).

59. F. Bassani, G. Iadonisi, and B. Preziosi. Electronic impurity levels in semiconductors. *Rep. Prog. Phys.* **37**, 1099–1210 (1974).
60. E. C. Stoner, Collective electron ferromagnetism. *Proc. R. Soc. Lond. A Math. Phys. Sci.* **165**, 372–414 (1997).
61. P. W. Anderson, Localized magnetic states in metals. *Phys. Rev.* **124**, 41–53 (1961).
62. G. C. Moore, M. K. Horton, E. Linscott, A. M. Ganose, M. Siron, D. D. O'Regan, and K. A. Persson, High-throughput determination of Hubbard $U$ and Hund $J$ values for transition metal oxides via the linear response formalism. *Phys. Rev. Mater.* **8**, 014409 (2024).
63. Y. J. Wang, Y. L. Tang, Y. L. Zhu, and X. L. Ma, Entangled polarizations in ferroelectrics: A focused review of polar topologies. *Acta Mater.* **243**, 118485 (2023).
64. S. Miyahara, S. Kusuta, and N. Furukawa, BCS theory on a flat band lattice. *Phys. C: Supercond.* **460–462**, 1145–1146 (2007).
65. J. M. D. Coey, Magnetism in $d^0$ oxides. *Nat. Mater.* **18**, 652–656 (2019).
66. R. Meng, L. da Costa Pereira, J. P. Locquet, V. Afanas'ev, G. Pourtois, and M. Houssa, Hole-doping induced ferromagnetism in 2D materials. *npj Comput. Mater.* **8**, 230 (2022).
67. R. Oja, M. Tyunina, L. Yao, T. Pinomaa, T. Kocourek, A. Dejneka, O. Stupakov, M. Jelinek, V. Trepakov, S. van Dijken, and R. M. Nieminen, $d^0$ ferromagnetic interface between nonmagnetic perovskites. *Phys. Rev. Lett.* **109**, 127207 (2012).
68. H. Lee, N. Campbell, J. Lee, T. J. Asel, T. R. Paudel, H. Zhou, J. W. Lee, B. Noesges, J. Seo, B. Park, L. J. Brillson, S. H. Oh, E. Y. Tsymbal, M. S. Rzchowski, and C. B. Eom. Direct observation of a two-dimensional hole gas at oxide interfaces. *Nat. Mater.* **17**, 231–236 (2018).
69. Q. Tong, F. Liu, J. Xiao, and W. Yao, Skyrmions in the Moiré of van der Waals 2D magnets. *Nano Lett.* **18**, 7194–7199 (2018).
70. F. Xiao, K. Chen, and Q. Tong, Magnetization textures in twisted bilayer Cr$X_3$ ($X$ = Br, I). *Phys. Rev. Res.* **3**, 013027 (2021).
71. K. Huang, E. Schwartz, D. Shao, A. A. Kovalev, and E. Y. Tsymbal, Magnetic antiskyrmions in two-dimensional van der Waals magnets engineered by layer stacking. *Phys. Rev. B* **109**, 024426 (2024).
72. F. Zheng, Magnetic skyrmion lattices in a novel 2D-twisted bilayer magnet. *Adv. Funct. Mater.* **33**, 2206923 (2023).
73. M. Akram and O. Erten, Skyrmions in twisted van der Waals magnets. *Phys. Rev. B* **103**, L140406 (2021).
74. T. Bömerich, L. Heinen, and A. Rosch, Skyrmion and tetarton lattices in twisted bilayer graphene. *Phys. Rev. B* **102**, 100408(R) (2020).
75. S. Ray and T. Das, Hierarchy of multi-order skyrmion phases in twisted magnetic bilayers *Phys. Rev. B* **104**, 014410 (2021).
76. M. Sheeraz, M. H. Jung, Y. K. Kim, N. J. Lee, S. Jeong, J. S. Choi, Y. J. Jo, S. Cho, I. W. Kim, Y. M. Kim, S. Kim, C. W. Ahn, S. M. Yang, H. Y. Jeong, and T. H. Kim. Freestanding oxide membranes for epitaxial ferroelectric heterojunctions. *ACS Nano* **17**, 13510–13521 (2023).
77. M.-S. Kim, K. Lee, R. Ishikawa, K. Song, N. A. Shahed, K.-T. Eom, M. S. Rzchowski, E. Y. Tsymbal, T. Mizoguchi, C.-B. Eom, and S.-Y. Choi, Twisted oxide membrane interface by local atomic registry design. *To be submitted* (2024).
78. K. Momma and F. Izumi, VESTA 3 for three-dimensional visualization of crystal, volumetric and morphology data. *J. Appl. Crystallogr.* **44**, 1272–1276 (2011).
79. G. Kresse and D. Joubert, From ultrasoft pseudopotentials to the projector augmented-wave method. *Phys. Rev. B* **59**, 1758–1775 (1999).
80. G. Kresse and J. Furthmüller, Efficient iterative schemes for ab initio total-energy calculations using a plane-wave basis set. *Phys. Rev. B* **54**, 11169–11186 (1996).
81. P. E. Blöchl, Projector augmented-wave method. *Phys. Rev. B* **50**, 17953–17979 (1994).
82. P. E. Blöchl, O. Jepsen, and O. K. Andersen, Improved tetrahedron method for Brillouin-zone integrations. *Phys. Rev. B* **49**, 16223–16233 (1994).
83. L. Bellaiche and D. Vanderbilt, Virtual crystal approximation revisited: Application to dielectric and piezoelectric properties of perovskites. *Phys. Rev. B* **61**, 7877 (2000).
84. T. Tsuneda and K. Hirao, Self-interaction corrections in density functional theory. *J. Chem. Phys.* **140**, 18A513 (2014).
85. T. A. Manz and N. G. Limas, Introducing DDEC6 atomic population analysis: Part 1. Charge partitioning theory and methodology. *RSC Adv.* **6**, 47771–47801 (2016).
86. T. A. Manz and N. G. Limas, Introducing DDEC6 atomic population analysis: Part 2. Computed results for a wide range of periodic and nonperiodic materials. *RSC Adv.* **6**, 45727–45747 (2016).
87. T. A. Manz, Introducing DDEC6 atomic population analysis: Part 3. Comprehensive method to compute bond orders. *RSC Adv.* **7**, 45552–45581 (2017).
88. T. A. Manz and N. G. Limas, Introducing DDEC6 atomic population analysis: Part 4. Efficient parallel computation of net atomic charges, atomic spin moments, bond orders, and more. *RSC Adv.* **8**, 2678–2707 (2018).
89. W. Tang, E. Sanville, and G. Henkelman, *J. Phys.: Condens. Matter* **21**, 084204 (2009).
90. T. Yusufaly, D. Vanderbilt, and S. Coh, Tight-binding formalism in the context of the PythTB package, http://physics.rutgers.edu/pythtb/formalism.html.